\pdfoutput=1
\documentclass[11pt]{article}

\usepackage[letterpaper,margin=1in]{geometry}
\usepackage[T1]{fontenc}
\usepackage[utf8]{inputenc}

\usepackage{amsmath,amssymb,amsfonts,amsthm,mathtools}
\usepackage{aliascnt}
\usepackage{microtype}
\usepackage{xcolor}
\usepackage{graphicx}
\usepackage{booktabs}
\usepackage{array}
\usepackage{enumitem}
\usepackage{parskip}
\usepackage{algorithm}
\usepackage{algorithmic}

\usepackage[square,numbers,semicolon]{natbib}
\usepackage{hyperref}
\hypersetup{
  colorlinks=true,
  linkcolor=blue,
  citecolor=blue,
  urlcolor=blue,
  filecolor=blue,
  anchorcolor=blue,
  pdftitle={The Movement Complexity of Sorting},
  pdfauthor={Sichen Wang}
}

\theoremstyle{plain}
\newtheorem{theorem}{Theorem}[section]
\newaliascnt{proposition}{theorem}
\newtheorem{proposition}[proposition]{Proposition}
\aliascntresetthe{proposition}
\newaliascnt{lemma}{theorem}
\newtheorem{lemma}[lemma]{Lemma}
\aliascntresetthe{lemma}
\newaliascnt{corollary}{theorem}

\aliascntresetthe{corollary}
\newaliascnt{conjecture}{theorem}
\newtheorem{conjecture}[conjecture]{Conjecture}
\aliascntresetthe{conjecture}
\newaliascnt{observation}{theorem}
\newtheorem{observation}[observation]{Observation}
\aliascntresetthe{observation}
\newaliascnt{fact}{theorem}
\newtheorem{fact}[fact]{Fact}
\aliascntresetthe{fact}

\theoremstyle{definition}
\newaliascnt{definition}{theorem}
\newtheorem{definition}[definition]{Definition}
\aliascntresetthe{definition}

\theoremstyle{remark}
\newaliascnt{remark}{theorem}
\newtheorem{remark}[remark]{Remark}
\aliascntresetthe{remark}
\newaliascnt{example}{theorem}

\aliascntresetthe{example}

\usepackage[capitalize,nameinlink]{cleveref}
\crefname{theorem}{Theorem}{Theorems}
\Crefname{theorem}{Theorem}{Theorems}
\crefname{proposition}{Proposition}{Propositions}
\Crefname{proposition}{Proposition}{Propositions}
\crefname{lemma}{Lemma}{Lemmas}
\Crefname{lemma}{Lemma}{Lemmas}
\crefname{corollary}{Corollary}{Corollaries}
\Crefname{corollary}{Corollary}{Corollaries}
\crefname{conjecture}{Conjecture}{Conjectures}
\Crefname{conjecture}{Conjecture}{Conjectures}
\crefname{observation}{Observation}{Observations}
\Crefname{observation}{Observation}{Observations}
\crefname{fact}{Fact}{Facts}
\Crefname{fact}{Fact}{Facts}

\crefname{algorithm}{Algorithm}{Algorithms}
\Crefname{algorithm}{Algorithm}{Algorithms}
\crefname{definition}{Definition}{Definitions}
\Crefname{definition}{Definition}{Definitions}
\crefname{remark}{Remark}{Remarks}
\Crefname{remark}{Remark}{Remarks}
\crefname{example}{Example}{Examples}
\Crefname{example}{Example}{Examples}
\crefname{section}{Section}{Sections}
\Crefname{section}{Section}{Sections}
\crefname{equation}{Equation}{Equations}
\Crefname{equation}{Equation}{Equations}
\crefname{table}{Table}{Tables}
\Crefname{table}{Table}{Tables}

\numberwithin{equation}{section}

\newcommand{\Mof}[2]{M_{#1}^{(#2)}}

\newcommand{\Tht}{\widetilde{\Theta}}

\newcommand{\eps}{\varepsilon}
\DeclareMathOperator{\rk}{rk}
\DeclareMathOperator{\dist}{dist}

\title{The Movement Complexity of Sorting}
\author{Sichen Wang\\[2pt]
  The Chinese University of Hong Kong, Shenzhen\\
  \texttt{sichenwang@cuhk.edu.cn}}
\date{}

\begin{document}
\maketitle

\begin{abstract}
The complexity of sorting is usually measured by the number of comparisons. This measure ignores the movement of
the read heads between comparisons, which is a real cost on many machines. We study this cost on its own. Keys lie on a
$d$-dimensional grid, two heads move over the grid and compare the keys beneath them, and the only cost is the total
movement of the heads. The algorithm plans its route in $r$ rounds and sees the outcomes of each round only after running
it. For every $d$ and $r$ we determine the expected movement needed to sort $n$ keys: it is $n^{\beta}$ up to polylogarithmic factors,
where $\beta=1+D^{r-1}/(D^{r}-(D-1)^{r})$ and $D=2d$, falling from $n^{2}$ at $r=1$ to $n^{1+1/(2d)}$ with unlimited
rounds. Our algorithm makes every round move about the same amount, and a matching lower bound holds against every algorithm.
\end{abstract}
\newpage

\section{Introduction}
\label{sec:intro}

The comparison model counts comparisons and nothing else, which for sorting gives the classical
$\Theta(n\log n)$. This count is not the whole cost of sorting. On general sequential models the product
of time and space for sorting is $\Omega(n^{2}/\log n)$~\cite{BorodinCook82}, later strengthened to
$\Omega(n^{2})$~\cite{Beame91}. On many machines the further cost is travel, because the two keys of a
comparison must first be brought within reach. On a Turing machine with one worktape and a two-way
read-only input tape, sorting an input of $N$ bits, made of strings of $3\log N$ bits each, takes
$\Omega(N^{3/2}/\sqrt{\log N})$ steps~\cite{MS86,DMS91}. This bound comes from the distance the heads must
travel: two heads on tapes of length $t$ need $\Omega(t^{3/2})$ steps to reach the pairs of tape positions
given by a well-scattered permutation~\cite{DMS91}.

Adaptivity has its own classical measure. In the parallel comparison model~\cite{Valiant75}, an
algorithm commits a set of comparisons in each round and learns the outcomes when the round ends, and
the exponents of the number of comparisons for sorting, selection, and merging in $r$ rounds are
known~\cite{Valiant75,BT83,HH82,AAV86,AA88,Pippenger87,BB90}. This theory charges nothing for moving
between comparisons, and the bounds on travel above say nothing about rounds. We ask: \emph{How much must
a processor travel to sort $n$ keys it cannot move, if it plans its route in $r$ rounds and learns the
outcomes of each round only after running it?}

We study the simplest model with this cost. The $n$ keys lie one at each position of the grid
$[m]^{d}$, where $n=m^{d}$, and they carry a hidden total
order. Two heads move over the grid in unit steps, and the algorithm learns about the order only by
comparing the two keys beneath the heads. It must output the list of positions in increasing order of
rank. The positions of the two heads form a point of the \emph{comparison grid} $[m]^{2d}$, and two keys are
compared when this point reaches their pair of positions, in either order. An
$r$-round algorithm commits a route at the start of each round, as a function of its coins and of the
outcomes of earlier rounds, and it receives the outcomes at all points of the route only after running it.
Comparisons, computation, and memory are free, and the only cost is the \emph{movement}, the total
$\ell_{1}$ distance the heads move. The algorithm must be correct on every input and for every outcome of
its coins, and $\Mof{r}{d}(n)$ is the least expected movement on a worst input. For $r=\infty$ the number
of rounds is free. \Cref{subsec:model} gives the precise definition. The grid is the basic metric space of polynomial
growth, and the proofs use little beyond the sizes of its balls and its partitions into cubes and boxes.

Two baselines bound the answer. For $r=1$, the oblivious case, the single route is fixed in advance and must compare
all $\binom{n}{2}$ pairs, so $\Mof{1}{d}(n)=\Theta(n^{2})$. For $r=\infty$, the fully adaptive case, a merge sort attains $O(n^{1+1/(2d)})$, and this
matches the lower bound of \Cref{prop:floor}.

\paragraph{Problem.}
For every dimension $d$ and every round budget $r$, determine $\Mof{r}{d}(n)$, the least
worst-case expected movement with which an $r$-round algorithm sorts $n$ keys on $[m]^{d}$.

\subsection{Our Results}
\label{subsec:results}

We determine the exponent for every $d$ and $r$.

\begin{theorem}[Movement Exponent]
\label{thm:main}
Fix $d\ge 1$ and $r\in\{1,2,\dots\}\cup\{\infty\}$, and let $D=2d$. For finite $r$ put
\[
  \beta_{r}^{(d)}=1+\frac{D^{\,r-1}}{D^{\,r}-(D-1)^{r}},
\]
so that $\beta_{1}^{(d)}=2$, and let $\beta_{\infty}^{(d)}=\lim_{r\to\infty}\beta_{r}^{(d)}=1+\tfrac{1}{2d}$.
Then $\Mof{r}{d}(n)=\Tht\bigl(n^{\beta_{r}^{(d)}}\bigr)$, and
$\Mof{r}{d}(n)=\Theta\bigl(n^{\beta_{r}^{(d)}}\bigr)$ whenever $d=1$, $r\le 2$, or $r=\infty$. The upper
bounds are attained by explicit polynomial-time algorithms, deterministic for $r=1$ and $r=\infty$ and
randomized otherwise.
\end{theorem}

The exponent decreases from
$2$ at $r=1$ to $1+\tfrac{1}{2d}$ as $r\to\infty$, each further round lowering
it by a geometrically shrinking amount (\Cref{rem:shape}), and for $r\ge 2$ it also decreases in $d$. The
number of heads, by contrast, does not affect it: two heads are what a comparison needs, and any fixed
number of heads changes the complexity by at most a constant factor (\Cref{prop:kheads}). We call the range
$d\ge 2$, $3\le r<\infty$ the \emph{interior}. The logarithmic factors hidden by $\Tht$ occur only there.
\Cref{tab:frontier} lists the exponent at small parameters, and \Cref{sec:highdim} ends with the proof of the theorem.

\begin{table}[!htb]
\centering
\caption{The movement exponent $\beta_{r}^{(d)}$. Every entry gives the exact order
$\Theta\bigl(n^{\beta_{r}^{(d)}}\bigr)$, except those marked $\dagger$, which lie in the interior. There
the order is known up to a polylogarithmic factor and conjectured exact (\Cref{conj:interior}).}
\label{tab:frontier}
\begin{tabular}{c|ccccc}
\toprule
 & $r=1$ & $r=2$ & $r=3$ & $r=4$ & $r=\infty$\\
\midrule
$d=1$ & $2$ & $\tfrac{5}{3}$ & $\tfrac{11}{7}$ & $\tfrac{23}{15}$ & $\tfrac{3}{2}$\\[2pt]
$d=2$ & $2$ & $\tfrac{11}{7}$ & $\tfrac{53}{37}^{\dagger}$ & $\tfrac{239}{175}^{\dagger}$ & $\tfrac{5}{4}$\\[2pt]
$d=3$ & $2$ & $\tfrac{17}{11}$ & $\tfrac{127}{91}^{\dagger}$ & $\tfrac{887}{671}^{\dagger}$ & $\tfrac{7}{6}$\\
\bottomrule
\end{tabular}
\end{table}

On the line the exponent is tied to selection rather than to sorting (\Cref{rem:selection}): $\beta_{r}^{(1)}=1+\tfrac12 s_{r}$, where $s_{r}=1+1/(2^{r}-1)$ is the
exponent of the worst-case number of comparisons that a deterministic algorithm needs to find the maximum of $n$ keys in $r$
rounds~\cite{Valiant75,HH82}, while the number of comparisons for sorting in $r$ rounds has exponent $1+1/r$~\cite{AAV86}.

\subsection{Overview}
\label{subsec:overview}

The cost of a round depends on how many pairs of positions it must visit and on how these pairs are spread. Call an
unordered pair of positions a \emph{cell}. If $Q$ cells are spread evenly over the $n^{2}$ points of the comparison grid,
their nearest-neighbor spacing is about $(n^{2}/Q)^{1/D}$, and a route through them moves about $Q\cdot(n^{2}/Q)^{1/D}$:
their number times their spacing, as for tours through random point sets~\cite{BHH59,Steele81,Yukich98}. The two baselines
are instances of this product. All $\binom{n}{2}$ cells have spacing $1$, which gives $n^{2}$. The $\Theta(n)$ cells of
consecutive ranks, which every algorithm must compare, have spacing about $n^{1/D}$ on a random input, which gives
$n^{1+1/(2d)}$.

With $r$ rounds, the early rounds can reduce the number of cells left for the final one, at the cost of their own movement.
Our algorithm makes every round move about the same amount. This gives one equation per round, and the solution of these equations
is the exponent of \Cref{thm:main}. The lower bound leads to the same equations, which is why the two exponents agree.

\paragraph{The Algorithm.}
The algorithm keeps \emph{segments}, sets of keys of consecutive ranks whose internal order is not
yet known, and refines them round by round. In round $1$ it chooses $n/b_{0}$ random positions, the \emph{splitters}, for a parameter $b_{0}$. One head stops
on each splitter in turn while the other sweeps the grid, so the movement is $n^{2}/b_{0}$, and the keys between two consecutive
splitters form a segment of about $b_{0}$ keys. The final round, the \emph{cleanup}, compares all pairs inside segments. If round $2$ is the final one, these
are about $nb_{0}$ cells, known in advance, and a serpentine route through them moves about their number times their spacing
(\Cref{subsec:routing}), as in the classical tours through given points~\cite{Few55,PB89}. Making the two movements equal gives the
exponent for $r=2$.

With more rounds, refinement rounds come in between. A refinement round selects \emph{pivots}, keys that are compared with every key of
their segment, and these cut the segments further. Selecting each key independently, at the rate the refinement needs, fails: the cells of the round are then spread over
the whole grid, and the route through them moves a polynomial factor more than the other rounds. So the round selects pivots in groups
of nearby positions. It cuts the grid into cubes of side $h$, selects each cube with probability $p$, and takes the keys of the selected
cubes as pivots. The cells then lie in the \emph{product boxes} $I\times J$ of a selected cube $J$ and another cube $I$, and the heads
go through these boxes one after another. If the segments entering the round have about $b$ keys, this moves about $p\,h\,nb$ for
$h=(n/b)^{1/D}$.

The refined segments can still have many keys, but they meet few cubes. List a key's segment in rank order, starting at the key: each new
cube is selected with probability $p$, so a pivot appears after $1/p$ cubes in expectation. Hence the \emph{dispersion}, the sum over the
keys of the number of cubes that the key's segment meets, has expectation about $n/p$, as for segments of $1/p$ keys. On the line, a bound
on the dispersion is all that the next round needs.

So the parameters of the algorithm (\Cref{alg:sampling}) are a decreasing sequence of rank scales $b_{j}=n^{a_{j}}$, $0\le j\le r-2$, the
\emph{schedule}, with finest scale $g=b_{r-2}$. On the line, round $j+1$ uses $p=1/b_{j}$ and $h=(n/b_{j-1})^{1/D}$. For $d\ge 2$ the
samples are more elaborate, as explained below, but the movement has the same exponent. Requiring every round to move the
same amount $n^{\beta}$ gives one equation per round:
\begin{align*}
  &\text{splitter:}\qquad & \beta &= 2-a_{0},\\
  &\text{refinement $j$:}\qquad & \beta &= 1+\frac{1-a_{j-1}}{D}+a_{j-1}-a_{j}
    \qquad(1\le j\le r-2),\\
  &\text{cleanup:}\qquad & \beta &= 1+a_{r-2}+\frac{1-a_{r-2}}{D}.
\end{align*}
The right-hand sides are the exponents of $n^{2}/b_{0}$, of $p\,h\,nb_{j-1}$, and of $ng\,(n/g)^{1/D}$, which is again a number times a
spacing. By \Cref{lem:schedule}, the unique solution has $\beta=\beta_{r}^{(d)}$.

For $d\ge 2$ the cleanup needs more. Visiting $Q$ cells in the product box of two cubes of side $h$ moves about $h\,Q^{1-1/D}$. On the line
this exponent is $\tfrac12$, and the sum of the $\sqrt{Q}$ over the boxes is at most the dispersion. For $d\ge 2$ the exponent is larger, and
the sum also depends on the sizes of the segments. So the cleanup needs small dispersion, for the tour between the product boxes, and a small
sum of squared segment sizes, for the visits inside them. Sampling cubes gives the first and sampling keys gives the second, so at $r=3$ the
refinement round draws both samples (\Cref{subsec:hd-r3}). For $r\ge 4$ a refinement round can be followed by another one, whose cost depends
on the dispersion of its input, counted with boxes of many sizes. So each refinement round draws one sample for each box size of a nested
partition of the grid, up to a threshold, and the analysis keeps an upper bound on the expected dispersion for each of these sizes
(\Cref{subsec:interior}). This costs a factor $\log n$ for the number of samples and a factor $\log n$ for the number of box sizes per sample.

\paragraph{The Lower Bound.}
By Yao's principle, it suffices to consider deterministic algorithms on the \emph{uniform input}, a uniformly random assignment of the ranks to
the positions~\cite{Yao77}. Call a cell \emph{co-blocked at scale $b$} if its two ranks lie in one block of $b$ consecutive ranks, the blocks
being cut from rank $1$ upward. There are $\Theta(ng)$ cells co-blocked at scale $g$, and the final round must visit every one of them whose two positions have not been compared in earlier rounds with another
position of their block. Take such a cell $\{u,v\}$. The earlier outcomes do not determine the order of $u$ and $v$, because a chain of outcomes from one to the other stays inside
the block and starts with a comparison of $u$ or $v$ with another position of the block. So these outcomes are consistent with two inputs that
give $u$ and $v$ adjacent ranks, in opposite orders. The final round runs the same route on both, and only the cell $\{u,v\}$ tells them apart, so
this route must visit it (\Cref{lem:pfinal}). On the uniform input a constant fraction of these cells are at distance about $(n/g)^{1/D}$ from all
others (\Cref{lem:scatter}). So if the earlier rounds compare only a small fraction of the positions inside their blocks, the final route moves
about $ng\,(n/g)^{1/D}$ (\Cref{fact:nn}), and the cleanup equation says that this is $n^{\beta}$.

It remains to show that rounds moving much less than $n^{\beta}$ compare few positions inside their blocks. The route of round $1$ is fixed before
any outcome is known, so each cell on it is co-blocked at scale $b_{0}$ with probability at most about $b_{0}/n$. A route of movement $n^{\beta}$
therefore visits about $n^{\beta}b_{0}/n$ such cells, and the splitter equation says that this number is $n$. So a much shorter route compares only a small fraction of the
positions inside their blocks. For $r=2$ we have $b_{0}=g$, and the proof is complete.

From round $2$ on, a route can use what the algorithm has learned to choose cells that are likely to be co-blocked, and this adaptivity is the hard part
of the lower bound. The proof handles it round by round, like the chronogram method for dynamic data-structure lower bounds~\cite{FS89}, with rounds in
place of epochs, and like round elimination in communication complexity~\cite{MNSW98}. For round $j+1$, two facts replace the estimate above.

The first bounds the number
of cells a route can visit. On the uniform input, with high probability, every route of movement $L$ visits at most about
$\bigl(1+L(b_{j-1}/n)^{1/D}\bigr)\log n$ cells co-blocked at scale $b_{j-1}$ (\Cref{lem:density}). This is a property of the input alone, so it holds for
adaptive routes.

The second bounds the probability that such a cell is co-blocked at the next scale. The proof gives the algorithm more information, which
can only help it: after round $j$, for $j\le r-2$, the index, or \emph{label}, of the block of each position at scale $b_{j-1}$ is revealed at no cost. Given
all labels and outcomes so far, the positions of a block that have not been compared with another position of the block still receive the ranks left to them in
uniformly random order, because exchanging their ranks changes no label and no outcome. So if at least half of the block is of this kind, a cell that is co-blocked at
scale $b_{j-1}$ and contains such a position is co-blocked at scale $b_{j}$ with conditional probability only $O(b_{j}/b_{j-1})$ (\Cref{lem:finehit}). Blocks with fewer
such positions are counted as entirely compared.

Multiplying the two bounds, a route of movement $n^{\beta}$ in round $j+1$ visits about
$n^{\beta}(b_{j-1}/n)^{1/D}\,b_{j}/b_{j-1}$ cells of this kind, up to a logarithmic factor, and the refinement equation says that this number is $n$. So here too a much shorter route
compares only a small fraction of the positions inside their blocks.

\paragraph{Removing the Logarithms.}
The factors $\log^{2}n$ in the upper bound and $\log n$ in the lower bound do not always occur. The factor $\log^{2}n$ comes only from the samples for many
box sizes, so the upper bound has no logarithmic factor for $d=1$ and for $r\le 3$. The factor $\log n$ comes only from \Cref{lem:density}, where the
largest number of co-blocked cells in one product box enters. For $r\le 2$ the route of the first round is fixed in advance, so this lemma is not needed. On
the line, \Cref{lem:sharp-density,lem:xi-bound} avoid the loss: they group nearby co-blocked cells into clusters, a route visits few cells of small clusters per
unit of movement, and the large clusters hold at most one cell in expectation. For $r=\infty$, a merge sort over a partition of the grid into boxes matches the
bound $n^{1+1/(2d)}$ given by the cells of consecutive ranks (\Cref{prop:floor,prop:floor-ub}).

\Cref{sec:prelims} defines the model and solves the equations, and \Cref{sec:toolkit} proves the geometric facts used by the lower bound.
\Cref{sec:lb} proves the lower bound, with the exact order on the line in \Cref{subsec:line-exact}, and
\Cref{sec:cascade,sec:highdim} prove the upper bounds. \Cref{sec:kheads} treats more than two
heads, and \Cref{sec:discussion} discusses the remaining gap.

\subsection{Related Work}
\label{subsec:related}

The models closest to ours attach costs to comparisons. Priced information charges every query and asks
for the cheapest correct computation~\cite{CFGKRS02}, and for sorting and selection the costs have been taken
structured or metric~\cite{GK01,GK05}. In the metric model a comparison costs the distance between its two keys.
There the costs of the comparisons add up independently, whereas here consecutive comparisons share the heads, so a
comparison costs only the move from the previous comparison. As a result, the cost of a round depends on how its
comparisons are distributed in the comparison grid, and not only on how many there are. \Cref{lem:batch,fact:nn}
give the upper and the lower bound for a given distribution. The benchmark also differs: these
works measure a competitive ratio against the cheapest proof of the order, while we determine the absolute cost on a
worst input. For sorting this ratio is polylogarithmic for metric costs~\cite{GK05}, and for general costs a
polylogarithmic ratio is known only in a bichromatic special case~\cite{GoswamiJacob24}. A separate line restricts which
comparisons are admissible at all, as in generalized sorting and the $1$-$\infty$ cost
model~\cite{HKK11,KN21,BR15}.

Sorting and selection in rounds come from Valiant's parallel comparison model~\cite{Valiant75}. They were studied
extensively there~\cite{BT83,HH82,AAV86,AA88,Pippenger87,BB90} and remain active~\cite{BMP19}. These results count
comparisons, and the place of a comparison plays no role in them. We keep the round structure and replace the cost by
movement, and on the line the movement exponent is determined by the exponent $s_{r}$ of \Cref{subsec:results}.

A universal tour orders $n$ points once and serves every subset in that order. In the plane, the space-filling curve
gives a competitive ratio of $\Theta(\log n)$~\cite{PB89,BG89}, and no universal tour has a constant
ratio~\cite{HKL06,Kravaris24}. On general metrics, sparse partitions~\cite{Filtser24} give polylogarithmic
ratios~\cite{JLNRS05,GHR06}. The a priori TSP asks the same question in expectation~\cite{Jaillet88}. These problems share
with ours the commitment of a route before it is known which visits matter. There the tour is fixed once and judged by
its ratio to the best tour for the subset, while our algorithm commits a new route every round and we determine the
absolute movement.

Travel has also been studied for element distinctness, on the Turing machine above~\cite{Petersen08} and on the one-tape
machine without an input tape~\cite{LopezOrtiz94,BBP03}. Other models charge for moving data. On a mesh-connected computer,
records travel along the links to their destinations~\cite{TK77}, and on a VLSI chip, the information that must cross the
chip gives area-time trade-offs~\cite{Thompson83}. Token swapping rearranges labeled tokens along the edges of a
graph and counts the swaps~\cite{Yamanaka15}. External-memory sorting transports records between memory levels, and
its classical lower bound treats the records as indivisible~\cite{AV88}. Under the network-coding conjecture of Li and
Li, the same bound holds without this assumption~\cite{LL04,FHLS19}. Here the keys stay in their positions, so the
lower bound needs neither assumption, and the comparisons it counts are forced by correctness alone (\Cref{rem:free}).

\section{Preliminaries}
\label{sec:prelims}

This section makes the model of \Cref{sec:intro} precise and solves the equations of \Cref{subsec:overview}, which
give the exponent and the rank scales used by both bounds.

\subsection{The Model}
\label{subsec:model}

\begin{definition}[The Movement Model]
\label{def:model}
Fix $d\ge 1$ and $m\ge 2$, let $n=m^{d}$, and write $[m]=\{1,\dots,m\}$.
\begin{enumerate}[label=\arabic*.]
  \item \emph{Input and output.} Every position $x\in[m]^{d}$ holds one key, and the keys carry a
        hidden total order: the input is a bijection $\pi\colon[m]^{d}\to[n]$, and
        $\rk_{\pi}(x)=\pi(x)$ is the rank of the key at $x$. The algorithm learns about $\pi$ only
        through the outcomes of item~3. The output is the list of positions in increasing order of
        rank.
  \item \emph{Configurations and routes.} Two heads occupy positions $x,y\in[m]^{d}$, and the pair
        $(x,y)$ is their \emph{configuration}. The configurations are the points of the
        \emph{comparison grid} $[m]^{D}$, where $D=2d$. A \emph{route} is a sequence of
        configurations $z_{0},\dots,z_{L}$ with $\lVert z_{t+1}-z_{t}\rVert_{1}=1$ for $0\le t<L$.
        The \emph{movement} of the route is $L$. Both heads start at the position $(1,\dots,1)$.
  \item \emph{Comparisons.} The \emph{outcome} at a configuration $(x,y)$ is whether
        $\rk_{\pi}(x)<\rk_{\pi}(y)$. Comparisons, computation, and memory are free.
  \item \emph{Rounds.} An \emph{$r$-round algorithm} proceeds in $r$ rounds. At the start of each
        round it commits, as a function of its random coins and of the outcomes received in
        earlier rounds, a route that starts at the current configuration. It then runs the route
        and receives the outcomes at all configurations on it. No choice made within a round
        depends on outcomes of that round.
  \item \emph{Correctness and cost.} The algorithm must output the correct order for every input
        and every outcome of its coins. We
        write $\mathrm{Move}(\mathrm{ALG},\pi)$ for the total movement of an algorithm
        $\mathrm{ALG}$ on input $\pi$, and
        \[
          \Mof{r}{d}(n)=\inf_{\mathrm{ALG}}\ \max_{\pi}\
            \mathbb{E}\bigl[\mathrm{Move}(\mathrm{ALG},\pi)\bigr],
        \]
        the infimum over $r$-round algorithms. For $r=\infty$, $\Mof{\infty}{d}(n)=\inf_{r}\Mof{r}{d}(n)$.
\end{enumerate}
\end{definition}

\begin{definition}[Cells]
\label{def:cell}
A \emph{cell} is an unordered pair $\{x,y\}$ of distinct positions. Its two \emph{orientations} are
the configurations $(x,y)$ and $(y,x)$. On cells, the metric $\dist$ is
\[
  \dist\bigl(\{x,y\},\{x',y'\}\bigr)
   =\min\bigl\{\lVert x-x'\rVert_{1}+\lVert y-y'\rVert_{1},\
                \lVert x-y'\rVert_{1}+\lVert y-x'\rVert_{1}\bigr\},
\]
and for a cell $w$ and a set $W$ of cells we put $\dist(w,W)=\min_{w'\in W}\dist(w,w')$, with
$\dist(w,\emptyset)=+\infty$.
\end{definition}

\begin{observation}
\label{obs:model}
\begin{enumerate}[label=\textup{(\alph*)}]
  \item A route of movement $L$ visits at most $L+1$ configurations.
  \item The two orientations of a cell carry the same information, and a configuration $(x,x)$
        carries none. On the configurations with $x\ne y$, the map
        $(x,y)\mapsto\{x,y\}$ is $1$-Lipschitz from $\ell_{1}$ to $\dist$. Hence deleting the
        configurations $(x,x)$ from a route of movement $L$ leaves a sequence of cells of
        $\dist$-length at most $L$. The number of cells within $\dist$-distance $\rho$ of a cell
        $\{x,y\}$ is at most the number of points of $[m]^{D}$ within $\ell_{1}$-distance $\rho$ of
        $(x,y)$.
        On the line, $(x,y)\mapsto(\min\{x,y\},\max\{x,y\})$ folds the configurations onto
        $\{x\le y\}$ and the cells onto $\{x<y\}$, and $\dist$ is the $\ell_{1}$ distance between
        the folded cells.
\end{enumerate}
\end{observation}

\begin{proof}
Part~(a) holds because the route consists of $z_{0},\dots,z_{L}$. For~(b), when $x\ne y$ the
outcome at $(y,x)$ is the negation of the outcome at $(x,y)$, and the outcome at $(x,x)$ is always
negative. By definition, $\dist(\{x,y\},\{x',y'\})$ is the smaller of the $\ell_{1}$ distances from
$(x,y)$ to the two orientations of $\{x',y'\}$, so the $\ell_{1}$ distance of two configurations is
at least the $\dist$ of their cells, which is the Lipschitz property. Since the swap
$(x,y)\mapsto(y,x)$ is an isometry of $\ell_{1}$, $\dist$ is a metric on cells. Between two
consecutive configurations that survive the deletion, the route moves at least their $\ell_{1}$
distance, so the cells have $\dist$-length at most $L$. A cell within $\dist$-distance $\rho$ of
$\{x,y\}$ has an orientation within $\ell_{1}$-distance $\rho$ of $(x,y)$, and distinct cells have no
orientation in common, which gives the count. On the line, if $x<y$ and $x'<y'$, then
$t\mapsto|t-y'|-|t-x'|$ is nonincreasing, and comparing its values at $x$ and at $y$ gives
$|x-x'|+|y-y'|\le|x-y'|+|y-x'|$. So the minimum in $\dist$ is the $\ell_{1}$ distance between
$(x,y)$ and $(x',y')$.
\end{proof}

Since the keys do not move, we identify each key with its position, so a set of keys is a set of
positions. When $\pi$ is fixed we write $\rk$ for $\rk_{\pi}$. The algorithm \emph{compares} $x$
and $y$ in round $t$ if the route of round $t$ visits an orientation of $\{x,y\}$. In the same
way, a route visits a set of cells if it visits an orientation of each, and a cell lies in a set of
configurations if one of its orientations does.

\paragraph{Notation.}
Throughout, $d$ and $r$ are fixed and asymptotic statements refer to $n\to\infty$. A subscript on
an asymptotic symbol lists the parameters other than $n$ on which the hidden constant depends, in
the order $d$, $r$, and, in \Cref{sec:kheads}, $k$, as in $O_{d}$ or $\Omega_{d,r}$. A symbol without a subscript hides an
absolute constant. \Cref{thm:main} carries no subscripts, and its constants depend on
$d$ and $r$. A tilde, as in $\Tht$, also hides factors polylogarithmic in $n$.
Logarithms are to base $2$. When the algorithm and the input are clear, we write $\mathrm{Move}$
for $\mathrm{Move}(\mathrm{ALG},\pi)$.

A \emph{box} is a product of intervals of integers, viewed as a set of lattice points. Its
\emph{side} along an axis is the number of its points on a line parallel to that axis, and a
\emph{cube} is a box whose sides are equal. A \emph{ball} is the set of grid points within a given
$\ell_{1}$ distance of a center. When we partition the line into
intervals of $h$ positions, or $[m]^{d}$ into cubes of side $H$, the last piece along each axis may
be shorter, which affects only constants.

\subsection{The Scale Schedule}
\label{subsec:schedule}

\begin{definition}[Scales and Blocks]
\label{def:scale}
A \emph{scale} is a number $b$ with $1\le b\le n$. For an integer scale $b$, the \emph{scale-$b$ blocks} are the sets of ranks
$\{(i-1)b+1,\dots,ib\}\cap[n]$ for $i=1,\dots,\lceil n/b\rceil$. Each holds $b$ ranks except
possibly the last, and $i$ is the \emph{label} of the $i$-th block. A position lies in a block if
its rank does. The scale $b$ is \emph{dyadic} if it is a power of two. Every block of a dyadic
scale lies in a block of each larger dyadic scale. For finite $r\ge 2$, a \emph{schedule} is a
sequence $b_{j}=n^{a_{j}}$, $0\le j\le r-2$, with exponents $1>a_{0}>a_{1}>\dots>a_{r-2}>0$. Its
finest term is $g=b_{r-2}$.
\end{definition}

For finite $r\ge 2$, both bounds use the schedule determined by the following system in the
unknowns $\beta,a_{0},\dots,a_{r-2}$. It has one equation for the splitter round of the algorithm,
one for each refinement round $j=1,\dots,r-2$, and one for the cleanup round. Each states that the
movement of that round is $n^{\beta}$, and \Cref{subsec:overview} explains the three expressions.
\begin{align}
  &\text{splitter:}\qquad & \beta &= 2-a_{0}, \label{eq:bal-split}\\
  &\text{refinement $j$:}\qquad & \beta &= 1+\frac{1-a_{j-1}}{D}+a_{j-1}-a_{j}
    \qquad(1\le j\le r-2), \label{eq:bal-ref}\\
  &\text{cleanup:}\qquad & \beta &= 1+a_{r-2}+\frac{1-a_{r-2}}{D}. \label{eq:bal-clean}
\end{align}

\begin{lemma}
\label{lem:schedule}
Let $r\ge 2$ and let $q=(D-1)/D$. The system of \Cref{eq:bal-split,eq:bal-ref,eq:bal-clean} has
the unique solution
\[
  \beta=\beta_{r}^{(d)}=1+\frac{1-q}{1-q^{\,r}},\qquad
  a_{j}=\frac{q^{\,j+1}-q^{\,r}}{1-q^{\,r}}\quad(0\le j\le r-2),
\]
and $1>a_{0}>a_{1}>\dots>a_{r-2}>0$, so the $a_{j}$ define a schedule. We write
$\delta=\beta_{r}^{(d)}-1=(1-q)/(1-q^{\,r})$, so $a_{0}=1-\delta$.
\end{lemma}

\begin{proof}
For a solution, write $\delta=\beta-1$ and $t_{j}=1-a_{j}$. \Cref{eq:bal-split} gives $t_{0}=\delta$.
Since $q=1-1/D$, \Cref{eq:bal-ref} reads $t_{j}=q\,t_{j-1}+\delta$, so
\[
  t_{j}=\delta\,(1+q+\dots+q^{\,j})\qquad(0\le j\le r-2),
\]
and \Cref{eq:bal-split,eq:bal-ref} determine every $a_{j}$ from $\delta$. \Cref{eq:bal-clean} reads
$\delta=1-q\,t_{r-2}$, that is, $\delta\,(1+q+\dots+q^{\,r-1})=1$, whose unique solution is
$\delta=(1-q)/(1-q^{\,r})$. Hence $\beta=\beta_{r}^{(d)}$, and $t_{j}=(1-q^{\,j+1})/(1-q^{\,r})$ gives
$a_{j}=(q^{\,j+1}-q^{\,r})/(1-q^{\,r})$. Finally,
$a_{j}-a_{j+1}=q^{\,j+1}(1-q)/(1-q^{\,r})>0$ for $0\le j\le r-3$,
$a_{r-2}=q^{\,r-1}(1-q)/(1-q^{\,r})>0$, and $a_{0}=1-\delta<1$.
\end{proof}

From now on, for given $d$ and finite $r\ge 2$, the exponents $a_{0},\dots,a_{r-2}$ are those of
\Cref{lem:schedule}, $b_{j}=n^{a_{j}}$ is their schedule, and $g=b_{r-2}$. The algorithms use the
$b_{j}$ as real parameters. In \Cref{sec:lb}, where blocks must nest, $b_{j}$ denotes the largest
power of two not exceeding $n^{a_{j}}$, which differs from $n^{a_{j}}$ by a factor less than $2$.

\begin{remark}
\label{rem:shape}
Since $(1-q)/(1-q^{\,r})=1/(1+q+\dots+q^{\,r-1})$, and the sum $1+q+\dots+q^{\,r-1}$ increases
strictly in $r$ and, for $r\ge 2$, in $q=1-1/(2d)$, the exponent $\beta_{r}^{(d)}$ decreases strictly in $r$
and, for $r\ge 2$, in $d$. It lies in $(1+\tfrac{1}{2d},\,2]$. Its distance to the fully adaptive
exponent satisfies
\[
  \beta_{r}^{(d)}-\Bigl(1+\frac{1}{2d}\Bigr)
   =\frac{q^{\,r}(1-q)}{1-q^{\,r}}.
\]
\end{remark}

\section{Geometry of the Comparison Grid}
\label{sec:toolkit}

This section proves three facts about the comparison grid that the lower bound uses. A route that
visits many cells far apart from one another has large movement (\Cref{fact:nn}). The other two facts
concern the \emph{uniform input}, a uniformly random bijection $\pi\colon[m]^{d}\to[n]$. With
probability at least $7/8$, at least half of the cells whose two positions lie in a common block of
ranks are far from all other such cells (\Cref{lem:scatter}). With probability at least $1-n^{-10}$,
every route, however it is chosen, visits few such cells per unit of movement (\Cref{lem:density}).

\begin{definition}[Co-Blocked Cells]
\label{def:coblocked}
Two positions are \emph{co-blocked} at an integer scale $b$ if their ranks lie in the same scale-$b$ block. A
cell is co-blocked at scale $b$ if its two positions are. We write $\mathcal{C}_{b}$ for the
set of cells co-blocked at scale $b$.
\end{definition}

The set $\mathcal{C}_{b}$ depends on the input, but its size does not. For integer $b$ with $2\le b\le n$ we have $\tfrac14 n(b-1)\le|\mathcal{C}_{b}|\le\tfrac12 n(b-1)$, and in particular
$|\mathcal{C}_{b}|\ge nb/8$. Indeed, $2|\mathcal{C}_{b}|$ is the sum over all positions $x$ of
the number of other positions co-blocked with $x$. Every position is co-blocked with at most $b-1$
others, which gives the upper bound. At least $n/2$ positions lie in blocks of exactly $b$ ranks,
since $b\lfloor n/b\rfloor\ge n/2$, and each of them is co-blocked with exactly $b-1$ others, which
gives the lower bound. On the uniform input, two fixed positions are co-blocked at scale $b$ with
probability at most $b/n$: given the rank of one, the rank of the other is uniform over the
remaining $n-1$ ranks, at most $b-1$ of which lie in its block, and $(b-1)/(n-1)\le b/n$.

\begin{fact}
\label{fact:nn}
Let $Y$ be a metric space with metric $\dist_{Y}$, and let $W\subseteq Y$ be finite with
$|W|\ge 2$. For $w\in W$ let $\nu_{W}(w)=\min_{w'\in W\setminus\{w\}}\dist_{Y}(w,w')$. Then every
finite sequence of points of $Y$ that contains every point of $W$ has length at least
$\tfrac12\sum_{w\in W}\nu_{W}(w)$, where the length of a sequence is the sum of the distances
between consecutive points.
\end{fact}

\begin{proof}
Keeping only the first occurrence of each point of $W$ gives a path through $W$ that, by the
triangle inequality, is no longer than the sequence. So the sequence is at least as long as a
minimum spanning tree of $W$. Charge each $w\in W$ to a shortest edge of this tree at $w$. That
edge has length at least $\nu_{W}(w)$, and each edge is charged at most twice, so
$\sum_{w\in W}\nu_{W}(w)$ is at most twice the length of the tree.
\end{proof}

We apply \Cref{fact:nn} to cells with the metric $\dist$. Let $\rho>0$, and let a route of movement
$L$ visit every cell of a set $W$ of at least two cells that are pairwise more than $\rho$ apart.
Deleting the configurations $(x,x)$ from the route leaves a sequence of cells of $\dist$-length at
most $L$ that contains every cell of $W$ (\Cref{obs:model}(b)). Since $\nu_{W}(w)>\rho$ for every
$w\in W$, \Cref{fact:nn} gives $L\ge\tfrac12|W|\rho$.

\begin{lemma}
\label{lem:scatter}
For every integer scale $b\ge 2$ there is a radius $\rho=\Theta_{d}\bigl((n/b)^{1/D}\bigr)$ such that, with
probability at least $7/8$ over the uniform input, at least half of the cells $w\in\mathcal{C}_{b}$
satisfy $\dist(w,\mathcal{C}_{b}\setminus\{w\})>\rho$.
\end{lemma}

\begin{proof}
We bound the expected number of pairs of cells of $\mathcal{C}_{b}$ that are close to each other
and apply Markov's inequality. Let $\rho=\eps_{0}(n/b)^{1/D}$, where $0<\eps_{0}\le 1$ depends only
on $d$ and is chosen below. Distinct cells are at distance at least $1$, so if $\rho<1$ every cell
satisfies the condition. Assume therefore that $\rho\ge 1$. Let $Z$ be the number of ordered pairs $(w,w')$ of
distinct cells of $\mathcal{C}_{b}$ with $\dist(w,w')\le\rho$. A cell $w\in\mathcal{C}_{b}$ with
$\dist(w,\mathcal{C}_{b}\setminus\{w\})\le\rho$ is the first entry of such a pair, so at most $Z$
cells violate the condition. It therefore suffices to show that
$\mathbb{E}Z\le\tfrac1{16}|\mathcal{C}_{b}|$: then Markov's inequality applied to $Z$ gives
$Z\le\tfrac12|\mathcal{C}_{b}|$ with probability at least $7/8$.

A ball of radius $\rho$ in $[m]^{d}$ contains at most $(2\rho+1)^{d}\le(3\rho)^{d}$ positions, and
by \Cref{obs:model}(b) at most $(3\rho)^{D}$ cells lie within distance $\rho$ of a given cell. Two
distinct cells share at most one position, so every pair counted by $Z$ falls into exactly one of
the following two cases.
\begin{itemize}
  \item The cells share a position: $w=\{u,v\}$ and $w'=\{u,v'\}$ with $v\ne v'$. Both lie in
        $\mathcal{C}_{b}$ only if the ranks of $u$, $v$, and $v'$ lie in one block. Given the rank of
        $u$, this has probability at most $\frac{b-1}{n-1}\cdot\frac{b-2}{n-2}\le(b/n)^{2}$. The
        condition $\dist(w,w')\le\rho$ requires $\lVert v-v'\rVert_{1}\le\rho$, or
        $\lVert u-v'\rVert_{1}+\lVert v-u\rVert_{1}\le\rho$, in which case both $v$ and $v'$ lie
        within distance $\rho$ of $u$. There are therefore at most $n^{2}(3\rho)^{d}+n(3\rho)^{2d}$
        such triples $(u,v,v')$, and these pairs contribute at most
        \[
          \Bigl(\frac{b}{n}\Bigr)^{2}\bigl(n^{2}(3\rho)^{d}+n(3\rho)^{2d}\bigr)
           =3^{d}\eps_{0}^{d}\,nb\Bigl(\frac{b}{n}\Bigr)^{1/2}+9^{d}\eps_{0}^{2d}\,b
           \le 2\cdot 9^{d}\eps_{0}^{d}\,nb .
        \]
  \item The cells have four distinct positions. There are at most $\tfrac12 n^{2}(3\rho)^{D}$ such
        pairs. For each of them both cells lie in $\mathcal{C}_{b}$ with probability $O(b^{2}/n^{2})$,
        because the first cell is co-blocked with probability at most $b/n$ and, given the ranks of
        its positions, the second with probability at most $(b-1)/(n-3)$. These pairs contribute
        \[
          O\Bigl(\frac{b^{2}}{n^{2}}\Bigr)\cdot n^{2}(3\rho)^{D}=O\bigl(9^{d}\eps_{0}^{2d}\,nb\bigr).
        \]
\end{itemize}
Since $|\mathcal{C}_{b}|\ge nb/8$ and $\eps_{0}\le 1$, there is a constant $c_{d}$ depending only
on $d$ such that the two contributions give $\mathbb{E}Z\le c_{d}\,\eps_{0}^{d}\,|\mathcal{C}_{b}|$.
Choosing $\eps_{0}$ with $c_{d}\eps_{0}^{d}\le\tfrac1{16}$ completes the proof.
\end{proof}

\begin{lemma}
\label{lem:density}
Fix a constant $\alpha<1$. There is a constant $c$ depending only on $d$ and $\alpha$ with the
following property. For a dyadic scale $b\le n^{\alpha}$, let $E_{b}$ be the event that every
route visits at most $c\bigl(1+L(b/n)^{1/D}\bigr)\log n$ cells of $\mathcal{C}_{b}$, where $L$ is the
movement of the route. Then, for sufficiently large $n$, with probability at least $1-n^{-10}$ over
the uniform input, $E_{b}$ holds for every dyadic scale $b\le n^{\alpha}$.
\end{lemma}

\begin{proof}
We partition the comparison grid into products of two cubes, show that with high probability none
of these products contains many cells of $\mathcal{C}_{b}$, and then cut a route into pieces that
each stay in few of them. Let $n$ be large enough for the estimates below, and fix a dyadic scale
$b\le n^{\alpha}$.

Let $s=\lceil(n/b)^{1/D}\rceil$ and partition $[m]^{d}$ into cubes of side $s$, clipped at the
boundary. Since $(n/b)^{1/D}\ge 1$, each cube contains at most $s^{d}\le V$ positions, where
$V=2^{d}(n/b)^{1/2}$. A \emph{product box} is a set $I\times J\subseteq[m]^{D}$ for two of these
cubes $I$ and $J$. There are at most $n^{2}$ product boxes. The \emph{load} of $I\times J$ is
$|\mathcal{C}_{b}\cap(I\times J)|$, the number of cells of $\mathcal{C}_{b}$ that lie in
$I\times J$. Cells that share a position are correlated, while disjoint cells are nearly independent, so we
show that a product box of large load contains either many disjoint cells or a block with many of its positions.
Let $\zeta=\lceil 30/(1-\alpha)\rceil$, $\mu=\lceil\max\{12\cdot 4^{d},\,13\log n\}\rceil$,
and $\Lambda_{0}=2\zeta\mu=O_{d,\alpha}(\log n)$. We show that a fixed product box $I\times J$ has
load at least $\Lambda_{0}$ with probability at most $2n^{-13}$. View the cells counted by the load
as the edges of a graph on $I\cup J$. Every edge joins two positions of one block. If every block
contains fewer than $\zeta$ positions of $I\cup J$, then every vertex has degree less than $\zeta$,
and a greedy matching, which discards fewer than $2\zeta$ edges for each edge it keeps, finds $\mu$
disjoint edges among any $\Lambda_{0}=2\zeta\mu$ edges. Hence, if the load is at least
$\Lambda_{0}$, one of the following two events occurs.
\begin{itemize}
  \item Some block contains at least $\zeta$ positions of $I\cup J$. The probability that $\zeta$
        given positions all lie in a given block is at most $(b/n)^{\zeta}$, and $I\cup J$ has at
        most $2V$ positions. A union bound over the at most $n$ blocks and the at most
        $(2V)^{\zeta}$ sets of $\zeta$ positions bounds the probability of this event by
        \[
          n\Bigl(\frac{2Vb}{n}\Bigr)^{\zeta}\le n\bigl(2^{d+1}n^{-(1-\alpha)/2}\bigr)^{\zeta}
           \le n^{-13},
        \]
        using $b\le n^{\alpha}$ and $(1-\alpha)\zeta/2\ge 15$.
  \item Some $\mu$ disjoint cells of $\mathcal{C}_{b}$ lie in $I\times J$. At most $V^{2}$ cells
        lie in $I\times J$, so there are at most $(V^{2})^{\mu}/\mu!$ sets of $\mu$ disjoint cells in
        $I\times J$. Given the ranks of the positions of the earlier cells of such a set, the next
        cell is co-blocked with probability at most $(b-1)/(n-2\mu)\le 2b/n$, so all its cells are
        co-blocked with probability at most $(2b/n)^{\mu}$. Since $V^{2}b/n=4^{d}$ and
        $\mu!\ge(\mu/3)^{\mu}$, a union bound over these sets bounds the probability of this event
        by
        \[
          \frac{(2\cdot 4^{d})^{\mu}}{\mu!}\le\Bigl(\frac{6\cdot 4^{d}}{\mu}\Bigr)^{\mu}
           \le 2^{-\mu}\le n^{-13}.
        \]
\end{itemize}
A union bound over the at most $\log n+1$ dyadic scales $b\le n^{\alpha}$ and the at most $n^{2}$
product boxes of each shows that, with probability at least $1-n^{-10}$, at every dyadic scale
$b\le n^{\alpha}$ every product box has load less than $\Lambda_{0}$.

On this event we bound the number of cells of $\mathcal{C}_{b}$ that a route visits. Cut a route
$z_{0},\dots,z_{L}$ into the pieces $z_{is},\dots,z_{\min\{(i+1)s,L\}}$ for
$0\le i\le\lfloor L/s\rfloor$. There are at most $1+L/s\le 1+L(b/n)^{1/D}$ of them. Every
configuration $(x',y')$ of a piece lies within $\ell_{1}$-distance $s$ of the first configuration
$(x,y)$ of the piece, so $x'$ lies in the cube of side $2s+1$ centered at $x$, and $y'$ in the cube
of side $2s+1$ centered at $y$. Along each axis, $2s+1$ consecutive integers meet at most three
intervals of the partition, so each of these two cubes meets at most $3^{d}$ cubes of the
partition, and the piece stays in at most $9^{d}$ product boxes. Every cell that the piece visits
lies in one of them, so the piece visits fewer than $9^{d}\Lambda_{0}$ cells of $\mathcal{C}_{b}$,
and the route visits fewer than $9^{d}\Lambda_{0}\bigl(1+L(b/n)^{1/D}\bigr)$. Since
$\Lambda_{0}=O_{d,\alpha}(\log n)$, this is the bound of $E_{b}$ for a suitable constant $c$
depending only on $d$ and $\alpha$.
\end{proof}

The logarithm in \Cref{lem:density} comes from the largest load over all product boxes. The
expected load of a fixed product box is at most $4^{d}$, since at most $4^{d}n/b$ cells lie in it,
each co-blocked with probability at most $b/n$.

\section{The Lower Bound}
\label{sec:lb}

This section proves the lower-bound half of \Cref{thm:main}.

\begin{theorem}[Lower Bound]
\label{thm:lb}
For every fixed $d\ge 1$ and $r\ge 1$, $\Mof{r}{d}(n)=\Omega_{d,r}\bigl(n^{\beta_{r}^{(d)}}/\log n\bigr)$.
\end{theorem}

The proof has the same plan for every finite $r\ge 2$. Correctness forces the final round to visit every cell that is
co-blocked at the finest scale and whose positions the earlier rounds have not compared inside their block
(\Cref{subsec:forcing}). Unless an earlier round already has movement about $n^{\beta_{r}^{(d)}}$, up to a logarithmic factor, many such cells
remain and lie far apart, so the final round has this movement. For $r\le 2$ this argument is short and loses no
logarithmic factor. \Cref{subsec:warmup} gives it, together with the bound $\Omega_{d}(n^{1+1/(2d)})$ for every $r$,
including $r=\infty$. For $r\ge 3$ the middle rounds are adaptive, and \Cref{subsec:badset} controls what they can
learn. \Cref{subsec:line-exact} removes the logarithmic factor on the line.

The lower bounds are proved on the uniform input. Fixing the coins of an $r$-round algorithm that is
correct on every input and for every outcome of its coins gives a deterministic $r$-round algorithm
that is correct on every input. Hence, by Yao's principle~\cite{Yao77}, a lower bound on the
expected movement on the uniform input that holds for every deterministic $r$-round algorithm
correct on every input is a lower bound on $\Mof{r}{d}(n)$. The lemmas of \Cref{subsec:forcing}
hold for every input, and the uniform input enters only when we count.

\subsection{The Augmented Model and Forced Comparisons}
\label{subsec:forcing}

For finite $r\ge 2$, let $\beta=\beta_{r}^{(d)}$. Recall from \Cref{subsec:schedule} that in this
section $b_{j}$ is the largest power of two not exceeding $n^{a_{j}}$, where $a_{0}>\dots>a_{r-2}$
are the exponents of \Cref{lem:schedule}, and that $g=b_{r-2}$. Thus
$n^{a_{j}}/2<b_{j}\le n^{a_{j}}$. The scales $b_{j}$ are powers of two, so every scale-$b_{j}$ block
lies in a scale-$b_{i}$ block for $i<j$. Since $a_{r-2}>0$ and $a_{j-1}>a_{j}$, for $n$ larger than a
constant depending on $d$ and $r$ we have $g\ge 2$ and $b_{j-1}\ge 2b_{j}$ for $1\le j\le r-2$, and we
assume this throughout the section.

A scale-$b$ block is \emph{full} if it holds $b$ ranks. Otherwise it is \emph{partial}. Every
$b_{j}$ divides $b_{0}$, so the positions whose rank lies in a partial block of some scale $b_{j}$
are exactly those of rank greater than $b_{0}\lfloor n/b_{0}\rfloor$, and there are fewer than
$b_{0}$ of them. The proofs below set these positions aside, because their counts need blocks of exactly $b$ ranks.

From round $2$ on, a route can use earlier outcomes to choose cells that are likely to be co-blocked, so the
probability $b/n$ no longer applies to the cells it visits. We therefore give the algorithm more information, which can
only help it, in a form that leaves, inside most blocks, the ranks of the positions not yet compared in the block
uniformly random (\Cref{lem:finehit}). In the \emph{augmented model}, after round $j$, for $1\le j\le r-2$, the label of the
scale-$b_{j-1}$ block of every position is revealed at no cost, and the route of each later round
may depend on the revealed labels. Thus the route of round $1$ depends on no outcome and no label,
and for $j\ge 1$ the route of round $j+1$ may depend on the outcomes of rounds $1,\dots,j$ and on the
labels revealed after them. The \emph{revealed scales} are $b_{0},\dots,b_{r-3}$, and the labels at the
finest scale $g$ are never revealed. An algorithm of the original model is an algorithm of the
augmented model that ignores the labels, so a lower bound in the augmented model is a lower bound in
the original one. Once the labels at a scale are revealed, a comparison of two positions in
different blocks of that scale has a known outcome, so all further information about the ranks comes
from comparisons inside these blocks.

We now show which comparisons correctness forces. Let $P_{t}$ be the partial order on positions generated by the relations $x<y$, where $\{x,y\}$
ranges over the cells compared in rounds $1,\dots,t$ and $\rk(x)<\rk(y)$. Revealed labels separate
only positions in different blocks, so they do not help to order two positions that have the same
labels:

\begin{lemma}
\label{lem:pfinal}
Let a deterministic $r$-round algorithm in the augmented model be correct on every input, and fix an
input. If two positions $u$ and $v$ are incomparable in $P_{r-1}$ and have the same label at every
revealed scale, then the route of round $r$ visits the cell $\{u,v\}$.
\end{lemma}

\begin{proof}
We construct two inputs that agree with the given input on all revealed labels and on all outcomes
of rounds $1,\dots,r-1$, but order $u$ and $v$ oppositely. Correctness then forces round $r$ to
compare $u$ and $v$.

If $r\le 2$, nothing is revealed, and we let $\lambda(x)=1$ for every position $x$. If $r\ge 3$, let
$\lambda(x)$ be the label of $x$ at the finest revealed scale $b_{r-3}$. The revealed partitions
nest, so $\lambda(x)$ determines the labels of $x$ at all revealed scales. In both cases
$\lambda(u)=\lambda(v)$. Let $P'$ be the partial order generated by $P_{r-1}$ and the relations $x<y$
for all positions $x,y$ with $\lambda(x)<\lambda(y)$. The given input extends $P'$, because blocks are intervals of ranks ordered
by their labels.

The positions $u$ and $v$ are incomparable in $P'$. Indeed, along a relation of $P_{r-1}$ the label
$\lambda$ does not decrease, because the two positions of a compared cell are ordered by rank and
every block is an interval of ranks. Along an added relation, $\lambda$ increases strictly. Hence
$\lambda$ does not decrease along a chain of $P'$ from $u$ to $v$, and since
$\lambda(u)=\lambda(v)$, it is constant on the chain. So the chain uses no added relation and lies
in $P_{r-1}$, which contradicts the incomparability of $u$ and $v$ in $P_{r-1}$. The same argument
rules out a chain from $v$ to $u$.

We now build the two inputs. List first the positions that lie below $u$ or $v$ in $P'$, in an
order that extends $P'$; then $u$ and $v$, in either order; then the remaining positions, in an
order that extends $P'$. No remaining position lies below a position of the first part, and since
$u$ and $v$ are incomparable in $P'$, neither of them lies below a position of the first part. So
both lists extend $P'$. Read as rank assignments, they are two inputs $\pi_{1}$ and $\pi_{2}$ that
differ by exchanging two adjacent ranks at $u$ and $v$. For $r\ge 3$, both have the given labels at
every revealed scale: in a linear extension of $P'$ the positions with $\lambda=i$ receive
consecutive ranks, and there are as many of them as the block with label $i$ has ranks, so they
receive exactly those ranks.

Finally we compare the executions. Since $\pi_{1}$ and $\pi_{2}$ extend $P_{r-1}$ and have the given
labels, induction over the rounds shows that on both inputs the algorithm runs the same routes in
rounds $1,\dots,r-1$ as on the given input and receives the same outcomes, so it also runs the same
route in round $r$. A comparison other than $\{u,v\}$ has the same outcome on $\pi_{1}$ and
$\pi_{2}$: if it involves neither $u$ nor $v$, its two ranks are unchanged, and if it involves $u$
or $v$ and a third position, the rank of the third position lies on the same side of the two
adjacent ranks. If the route of round $r$ did not visit $\{u,v\}$, all outcomes on $\pi_{1}$ and
$\pi_{2}$ would agree, the algorithm would output the same list on both inputs, and it would be
wrong on one of them.
\end{proof}

Two positions of one block can be ordered only through a comparison inside the block:

\begin{lemma}
\label{lem:prigid}
Let $P$ be the partial order generated, in the same way as $P_{t}$, by the outcomes of a set of
comparisons. If $u<_{P}v$ and $u$ and $v$ lie in a common scale-$b$ block, then one of these
comparisons involves $u$ and another position of this block.
\end{lemma}

\begin{proof}
Write $u<_{P}v$ as a chain $u=x_{0},x_{1},\dots,x_{i}=v$ in which every two consecutive positions
form a compared cell and have increasing ranks. The ranks along the chain lie between $\rk(u)$ and
$\rk(v)$, hence in the block. So the first cell of the chain, $\{u,x_{1}\}$, is a comparison of $u$
with another position of the block.
\end{proof}

Together, \Cref{lem:pfinal,lem:prigid} show that round $r$ must visit every cell co-blocked at scale
$g$ whose two positions have not been compared, in rounds $1,\dots,r-1$, with other positions of
their scale-$g$ block. Indeed, by \Cref{lem:prigid} the two positions are incomparable in $P_{r-1}$,
and since every revealed scale is coarser than $g$, they have the same revealed labels.

\subsection{Two Rounds and the Fully Adaptive Bound}
\label{subsec:warmup}

The proof for $r=2$ has four stages, which the proof for $r\ge 3$ in \Cref{subsec:badset} follows. The route of
round $1$ is fixed before any outcome is known, so there is no adaptivity to control, \Cref{lem:density} is not needed,
and no logarithmic factor is lost.

\begin{proposition}
\label{prop:endpoints}
For every fixed $d$ and every $r\le 2$, $\Mof{r}{d}(n)=\Omega_{d}\bigl(n^{\beta_{r}^{(d)}}\bigr)$.
\end{proposition}

\begin{proof}
By Yao's principle, fix a deterministic $r$-round algorithm that is correct on every input and let
$\pi$ be uniform. We show that $\mathbb{E}[\mathrm{Move}]=\Omega_{d}\bigl(n^{\beta_{r}^{(d)}}\bigr)$.

For $r=1$, the order $P_{0}$ is empty and no label is revealed, so by \Cref{lem:pfinal} the route
visits every cell. By \Cref{obs:model}(a), $\mathrm{Move}\ge\binom{n}{2}-1=\Omega(n^{2})$, and
$\beta_{1}^{(d)}=2$.

Now let $r=2$, so that $g=b_{0}$ and, by \Cref{eq:bal-split}, $\beta=2-a_{0}$. Write $L_{1}$ and
$L_{2}$ for the movements of the two routes. No outcome is known before round $1$, so the route of
round $1$ and $L_{1}$ do not depend on $\pi$. Let $\eps=1/192$. If $L_{1}\ge\eps n^{\beta}$, then
$\mathrm{Move}\ge\eps n^{\beta}$ and there is nothing to prove, so assume $L_{1}<\eps n^{\beta}$.
Call a position \emph{clean} if its rank lies in a full scale-$g$ block and it lies in no cell that
round $1$ visits and that is co-blocked at scale $g$.

\emph{Exposure.} The route of round $1$ visits at most $L_{1}+1$ cells (\Cref{obs:model}(a)), each
co-blocked at scale $g$ with probability at most $g/n$, and fewer than $g$ positions lie in a partial
block. So the expected number of positions that are not clean is less than
$g+2(L_{1}+1)g/n\le n^{a_{0}}+2\eps n^{\beta+a_{0}-1}+2$. Since $\beta+a_{0}-1=1$, this is
$n/96+o(n)$, which is at most $n/64$ for large $n$.

\emph{Survivors.} Every position lies in fewer than $g$ cells of $\mathcal{C}_{g}$, so the expected
number of cells of $\mathcal{C}_{g}$ with a position that is not clean is at most
$gn/64\le|\mathcal{C}_{g}|/8$. By Markov's inequality, with probability at least $1/2$ at most
$|\mathcal{C}_{g}|/4$ cells of $\mathcal{C}_{g}$ have such a position. By \Cref{lem:scatter} at
scale $g$, with probability at least $7/8$ at least half of the cells $w\in\mathcal{C}_{g}$ satisfy
$\dist(w,\mathcal{C}_{g}\setminus\{w\})>\rho$, where $\rho=\Theta_{d}\bigl((n/g)^{1/D}\bigr)$. On the
intersection of the two events, which has probability at least $1-\tfrac12-\tfrac18=\tfrac38$, let
$\mathcal{H}$ be the set of cells of $\mathcal{C}_{g}$ that have two clean positions and satisfy this
condition. Then $|\mathcal{H}|\ge|\mathcal{C}_{g}|/4\ge ng/32$, and any two cells of $\mathcal{H}$
are more than $\rho$ apart.

\emph{Forcing.} Let $\{u,v\}\in\mathcal{H}$. Since $u$ and $v$ are clean, neither has been compared in
round $1$ with another position of its scale-$g$ block, so by \Cref{lem:prigid} they are
incomparable in $P_{1}$. No label is revealed, so by \Cref{lem:pfinal} the route of round $2$ visits
$\{u,v\}$.

\emph{Cost of the Final Round.} Since $|\mathcal{H}|\ge 2$, \Cref{obs:model}(b) and \Cref{fact:nn}
give
\[
  L_{2}\ \ge\ \tfrac12|\mathcal{H}|\rho
   \ =\ \Omega_{d}\bigl(ng\,(n/g)^{1/D}\bigr)
   \ =\ \Omega_{d}\bigl(n^{1+a_{0}+(1-a_{0})/D}\bigr)
   \ =\ \Omega_{d}(n^{\beta}),
\]
where the second step uses $g>n^{a_{0}}/2$ and the last is \Cref{eq:bal-clean} for $r=2$. So, when
$L_{1}<\eps n^{\beta}$, we have $L_{2}=\Omega_{d}(n^{\beta})$ on an event of probability at least $3/8$, and hence
$\mathbb{E}[L_{2}]=\Omega_{d}(n^{\beta})$. In both cases
$\mathbb{E}[\mathrm{Move}]=\Omega_{d}(n^{\beta})$.
\end{proof}

The next bound holds for every number of rounds and needs no schedule. It uses the cells of
$\mathcal{C}_{2}$, which join the positions of ranks $2i-1$ and $2i$.

\begin{proposition}
\label{prop:floor}
For every fixed $d$ and every $r$, including $r=\infty$,
$\Mof{r}{d}(n)=\Omega_{d}\bigl(n^{1+1/(2d)}\bigr)$.
\end{proposition}

\begin{proof}
Let $r$ be finite. By Yao's principle, fix a deterministic $r$-round algorithm that is correct on
every input and let $\pi$ be uniform. Every cell $\{u,v\}\in\mathcal{C}_{2}$ joins the positions of ranks $2i-1$ and $2i$ for some $i$, and every
correct algorithm must compare it. Indeed, if the algorithm never compares $u$ and $v$, then
exchanging their two ranks changes the outcome of no comparison that it makes. So, by induction
over the rounds, it runs the same routes on both inputs and gives the same output, which is wrong on
one of them. By \Cref{lem:scatter} at scale $2$, with probability at least $7/8$ at least half of
the $\lfloor n/2\rfloor$ cells $w\in\mathcal{C}_{2}$ satisfy $\dist(w,\mathcal{C}_{2}\setminus\{w\})>\rho$,
where $\rho=\Theta_{d}(n^{1/(2d)})$, and these cells are pairwise more than $\rho$ apart. The routes of
all rounds, joined end to end, form one route of movement $\mathrm{Move}$ that visits all of them,
so by \Cref{obs:model}(b) and \Cref{fact:nn},
$\mathrm{Move}\ge\tfrac12\cdot\tfrac12\lfloor n/2\rfloor\cdot\rho=\Omega_{d}(n^{1+1/(2d)})$ on this
event. Hence $\mathbb{E}[\mathrm{Move}]=\Omega_{d}(n^{1+1/(2d)})$. The constant depends only on $d$,
so the bound also holds for $\Mof{\infty}{d}(n)=\inf_{r}\Mof{r}{d}(n)$.
\end{proof}

\subsection{Adaptive Rounds}
\label{subsec:badset}

Throughout \Cref{subsec:badset,subsec:line-exact}, let $r\ge 3$, fix a deterministic
$r$-round algorithm in the augmented model that is correct on every input, let $\pi$ be uniform, and
write $L_{t}$ for the movement of its route in round $t$. Since rounds $2,\dots,r-1$ are adaptive, we keep
track of what the first $r-1$ rounds can have learned. The bad set contains, scale by scale, the
positions that these rounds compare inside their blocks. It also contains all positions of
every block in which more than half of the positions are already bad, so that every other block keeps at
least half of its positions outside the bad set, which \Cref{lem:finehit} uses.

\begin{definition}[The Bad Set]
\label{def:badset}
The \emph{bad set} is built in levels $j=0,1,\dots,r-2$, and level $j$ concerns round $j+1$. It starts from the set $\mathcal{A}_{-1}$ of positions whose rank lies in a partial block
of some scale $b_{i}$. Level $j$ has two steps.
\begin{enumerate}[label=\textup{(\alph*)}]
  \item For $j\ge 1$, a scale-$b_{j-1}$ block is \emph{heavy at level $j$} if more than half of its
        positions lie in $\mathcal{A}_{j-1}$. Let $\mathcal{A}_{j}^{-}$ consist of
        $\mathcal{A}_{j-1}$ and all positions of the blocks that are heavy at level $j$. Let
        $\mathcal{A}_{0}^{-}=\mathcal{A}_{-1}$.
  \item Let $X_{j}$ be the set of positions outside $\mathcal{A}_{j}^{-}$ that lie in a cell visited
        in round $j+1$ and co-blocked at scale $b_{j}$, and let
        $\mathcal{A}_{j}=\mathcal{A}_{j}^{-}\cup X_{j}$.
\end{enumerate}
The bad set is $\mathcal{A}=\mathcal{A}_{r-2}$. A position is \emph{bad} if it lies in
$\mathcal{A}$, and \emph{clean} otherwise. It is \emph{clean at level $j$} if it lies outside
$\mathcal{A}_{j}^{-}$.
\end{definition}

\begin{observation}
\label{obs:clean}
\begin{enumerate}[label=\textup{(\alph*)}]
  \item For $1\le j\le r-2$, a position clean at level $j$ lies in a full scale-$b_{j-1}$ block that
        is not heavy at level $j$, and in rounds $1,\dots,j$ it has not been compared with another
        position of this block.
  \item A clean position has not been compared, in rounds $1,\dots,r-1$, with another position of
        its scale-$g$ block.
\end{enumerate}
\end{observation}

\begin{proof}
For every $i$, both positions of a cell that round $i+1$ visits and that is co-blocked at scale
$b_{i}$ lie in $\mathcal{A}_{i}$. Let $x$ be clean at level $j$, and suppose that in round
$i+1\le j$ it was compared with another position $y$ of its scale-$b_{j-1}$ block. The cell
$\{x,y\}$ is co-blocked at scale $b_{j-1}$, hence at the coarser scale $b_{i}$, so
$x\in\mathcal{A}_{i}\subseteq\mathcal{A}_{j}^{-}$, a contradiction. The block of $x$ is full because
$\mathcal{A}_{-1}\subseteq\mathcal{A}_{j}^{-}$, and it is not heavy because otherwise
$x\in\mathcal{A}_{j}^{-}$. This proves~(a). For~(b), let $x$ be clean, and suppose that in round
$i+1\le r-1$ it was compared with another position $y$ of its scale-$g$ block. The cell $\{x,y\}$ is
co-blocked at scale $g$, hence at scale $b_{i}$, so $x\in\mathcal{A}_{i}\subseteq\mathcal{A}$, a
contradiction.
\end{proof}

To keep the bad set small, we bound $\mathbb{E}|X_{j}|$ for $1\le j\le r-2$ by combining two facts. A position enters $X_{j}$
only if it is clean at level $j$ and lies in a cell that round $j+1$ visits and that is co-blocked
at scale $b_{j}$, and hence also at scale $b_{j-1}$. \Cref{lem:finehit} below shows that a cell
that is co-blocked at scale $b_{j-1}$ and contains a position clean at level $j$ is co-blocked at scale $b_{j}$
with conditional probability $O(b_{j}/b_{j-1})$, given what the algorithm knows before round
$j+1$, and \Cref{lem:density} shows that a short route visits few cells co-blocked at scale
$b_{j-1}$. \Cref{lem:exposure} combines the two.

For $1\le j\le r-2$, let $\mathcal{F}_{j}$ be the $\sigma$-field generated by the labels at scales
$b_{0},\dots,b_{j-1}$ and the outcomes of rounds $1,\dots,j$. In words, $\mathcal{F}_{j}$ is what
the algorithm knows when it commits the route of round $j+1$. An \emph{atom} of $\mathcal{F}_{j}$ is
one possible value of these labels and outcomes. Given an atom, that route is fixed. The lemma below
conditions on the positions clean at level $j$, so they must be determined by $\mathcal{F}_{j}$. Indeed, by induction on $i$, the sets $\mathcal{A}_{i-1}$ and
$\mathcal{A}_{i}^{-}$ are determined by $\mathcal{F}_{j}$ for every $i\le j$. The set
$\mathcal{A}_{-1}$ is determined by the labels at scale $b_{0}$. Whether a block is heavy at level
$i$ depends only on $\mathcal{A}_{i-1}$ and the labels at scale $b_{i-1}$. For $i<j$, the set
$X_{i}$ depends only on $\mathcal{A}_{i}^{-}$, the route of round $i+1$, and the labels at scale
$b_{i}$, and the last two are determined by $\mathcal{F}_{i+1}\subseteq\mathcal{F}_{j}$.

\begin{lemma}
\label{lem:finehit}
Let $1\le j\le r-2$, and let $\{u,v\}$ be a cell. On the event that
$\{u,v\}$ is co-blocked at scale $b_{j-1}$ and $u$ is clean at level $j$,
\[
  \Pr\bigl[\{u,v\}\text{ is co-blocked at scale }b_{j}\bigm|\mathcal{F}_{j}\bigr]
   \le\frac{4b_{j}}{b_{j-1}}.
\]
\end{lemma}

\begin{proof}
Fix an atom of $\mathcal{F}_{j}$ on which $\{u,v\}$ is co-blocked at scale $b_{j-1}$ and $u$ is
clean at level $j$, and let $B$ be the scale-$b_{j-1}$ block of $u$. By \Cref{obs:clean}(a), $B$ is
full and not heavy at level $j$. Since $\mathcal{A}_{j}^{-}$ adds to $\mathcal{A}_{j-1}$ only
positions of heavy blocks, at most half of the $b_{j-1}$ positions of $B$ lie in
$\mathcal{A}_{j}^{-}$, so the set $Y_{B}$ of positions of $B$ that are clean at level $j$ has at
least $b_{j-1}/2$ elements. Condition further on the ranks of all positions outside $Y_{B}$. The
ranks of $B$ not assigned to these positions are the \emph{free slots}. The positions of $Y_{B}$
occupy them.

We show that $\rk$ restricted to $Y_{B}$ is a uniformly random bijection onto the free slots. Since
$\pi$ is uniform, it suffices to show that permuting the ranks of $Y_{B}$ among the free slots does
not change the atom of $\mathcal{F}_{j}$. Such a permutation keeps every rank inside $B$, so it
changes no label at scales $b_{0},\dots,b_{j-1}$. By \Cref{obs:clean}(a), every comparison of
rounds $1,\dots,j$ that involves a position of $Y_{B}$ has its other position outside $B$. The rank
of that position lies above all ranks of $B$ or below all of them, so the outcome does not change.
The outcome of a comparison that involves no position of $Y_{B}$ does not change either. By induction
over the rounds, the routes and outcomes of rounds $1,\dots,j$ do not change, so the atom of
$\mathcal{F}_{j}$ does not change.

It remains to count. Since $\{u,v\}$ is co-blocked at scale $b_{j-1}$, $v$ lies in $B$. If
$v\notin Y_{B}$, the rank of $v$ is fixed, and $\rk(u)$ is uniform over at least $b_{j-1}/2$ free
slots, of which at most $b_{j}$ lie in the scale-$b_{j}$ block of $\rk(v)$, so the probability is at
most $2b_{j}/b_{j-1}$. If $v\in Y_{B}$, condition also on $\rk(v)$. Then $\rk(u)$ is uniform over the
remaining at least $b_{j-1}/2-1$ free slots, of which at most $b_{j}$ lie in the scale-$b_{j}$ block
of $\rk(v)$, so the probability is at most $b_{j}/(b_{j-1}/2-1)\le 4b_{j}/b_{j-1}$, because
$b_{j-1}\ge 2b_{j}\ge 4$. These bounds hold for every value of the further conditioning and every
such atom of $\mathcal{F}_{j}$, which proves the lemma.
\end{proof}

\begin{lemma}
\label{lem:exposure}
We have $\mathbb{E}|X_{0}|\le 2(L_{1}+1)\,b_{0}/n$, and for every
$1\le j\le r-2$,
\[
  \mathbb{E}|X_{j}|=O_{d,r}\Bigl(\frac{b_{j}}{b_{j-1}}
    \Bigl(1+\mathbb{E}[L_{j+1}]\Bigl(\frac{b_{j-1}}{n}\Bigr)^{1/D}\Bigr)\log n\Bigr).
\]
\end{lemma}

\begin{proof}
No outcome and no label is known before round $1$, so its route and $L_{1}$ do not depend on $\pi$,
and by \Cref{obs:model}(a) the route visits at most $L_{1}+1$ cells. Each of them is co-blocked at scale
$b_{0}$ with probability at most $b_{0}/n$ and contributes at most two positions to $X_{0}$, which
gives the bound on $\mathbb{E}|X_{0}|$.

Now let $1\le j\le r-2$, and fix an atom of $\mathcal{F}_{j}$. The route of round $j+1$ is then
fixed, and so is the set $K_{j}$ of the cells that it visits, that are co-blocked at scale
$b_{j-1}$, and that contain a position clean at level $j$. Every position of $X_{j}$ is clean at
level $j$ and lies in a cell visited in round $j+1$ and co-blocked at scale $b_{j}$, hence also at
scale $b_{j-1}$, so it lies in a cell of $K_{j}$ that is co-blocked at scale $b_{j}$. Applying
\Cref{lem:finehit} to each cell of $K_{j}$, with $u$ a position clean at level $j$, gives
\[
  \mathbb{E}\bigl[|X_{j}|\bigm|\mathcal{F}_{j}\bigr]\le 2\cdot|K_{j}|\cdot\frac{4b_{j}}{b_{j-1}}.
\]
We bound $|K_{j}|$ by \Cref{lem:density} with $\alpha=a_{0}$, which applies at scale $b_{j-1}$
because $a_{0}<1$ and $b_{j-1}\le b_{0}\le n^{a_{0}}$. Its constant $c$ then depends only on $d$
and $r$. The event $E_{b_{j-1}}$ depends only on $\mathcal{C}_{b_{j-1}}$, which the labels at scale
$b_{j-1}$ determine, so it is determined by $\mathcal{F}_{j}$. On $E_{b_{j-1}}$ we have
$|K_{j}|\le c\bigl(1+L_{j+1}(b_{j-1}/n)^{1/D}\bigr)\log n$, so taking expectations gives
\[
  \mathbb{E}\bigl[|X_{j}|\,\mathbf{1}_{E_{b_{j-1}}}\bigr]
   \le\frac{8c\,b_{j}}{b_{j-1}}\Bigl(1+\mathbb{E}[L_{j+1}]\Bigl(\frac{b_{j-1}}{n}\Bigr)^{1/D}\Bigr)
     \log n.
\]
Off $E_{b_{j-1}}$ we use $|X_{j}|\le n$. The complement of $E_{b_{j-1}}$ has probability at most
$n^{-10}$ for large $n$, so it contributes at most $n^{-9}$, which is at most $b_{j}/b_{j-1}$ because
$b_{j}\ge 1$ and $b_{j-1}\le n$. Adding the two parts proves the bound on $\mathbb{E}|X_{j}|$.
\end{proof}

The first stage of the proof of \Cref{thm:lb} uses \Cref{lem:exposure} together with the following observation.

\begin{observation}
\label{obs:badset-size}
For $0\le j\le r-2$, $|\mathcal{A}_{j}|\le 3|\mathcal{A}_{j-1}|+|X_{j}|$. Consequently
$\mathbb{E}|\mathcal{A}|\le 3^{r-1}\bigl(|\mathcal{A}_{-1}|+\max_{0\le j\le r-2}\mathbb{E}|X_{j}|\bigr)$.
\end{observation}

\begin{proof}
At level $j\ge 1$, a full heavy block holds more than $b_{j-1}/2$ positions of $\mathcal{A}_{j-1}$,
so at most $2|\mathcal{A}_{j-1}|/b_{j-1}$ full blocks are heavy, and together they hold at most
$2|\mathcal{A}_{j-1}|$ positions. The positions of a partial block lie in
$\mathcal{A}_{-1}\subseteq\mathcal{A}_{j-1}$ already. Hence $|\mathcal{A}_{j}^{-}|\le
3|\mathcal{A}_{j-1}|$, also for $j=0$, and the first claim follows from
$\mathcal{A}_{j}=\mathcal{A}_{j}^{-}\cup X_{j}$. Iterating it over the $r-1$ levels gives
$|\mathcal{A}|\le 3^{r-1}|\mathcal{A}_{-1}|+\sum_{j=0}^{r-2}3^{r-2-j}|X_{j}|$, and the second claim
follows by taking expectations, because $\sum_{j=0}^{r-2}3^{r-2-j}<3^{r-1}$.
\end{proof}

\begin{proof}[Proof of \Cref{thm:lb}]
For $r\le 2$ the theorem is \Cref{prop:endpoints}, so let $r\ge 3$. By Yao's principle, and because
an algorithm of the original model is an algorithm of the augmented model that ignores the labels,
it suffices to show that the algorithm fixed in \Cref{subsec:badset} has
$\mathbb{E}[\mathrm{Move}]=\Omega_{d,r}(n^{\beta}/\log n)$. The proof has the four stages of the
proof of \Cref{prop:endpoints}. The new difficulty is that rounds $2,\dots,r-1$ are adaptive. We
first show that the bad set is small, then find many cells of $\mathcal{C}_{g}$ with clean positions
that are pairwise far apart, show that round $r$ visits all of them, and bound the movement of a
route that does so.

Recall that $L_{t}$ is the movement of the route of round $t$, so that
$\mathrm{Move}=L_{1}+\dots+L_{r}$. Let $T=\eps n^{\beta}/\log n$, where $\eps>0$ depends only on
$d$ and $r$ and is chosen below. If $\mathbb{E}[L_{t}]\ge T/100$ for
some $t\le r-1$, then $\mathbb{E}[\mathrm{Move}]\ge T/100$, and we call this Case~1. Otherwise
$\mathbb{E}[L_{t}]<T/100$ for every $t\le r-1$. This is Case~2, which we assume until the last
paragraph.

\emph{Exposure.} We show that $\mathbb{E}|\mathcal{A}|\le n/64$ for large $n$. By
\Cref{lem:exposure} and Case~2, $\mathbb{E}|X_{0}|\le 2(T/100+1)b_{0}/n$. Since
$b_{0}\le n^{a_{0}}$ and, by \Cref{eq:bal-split}, $\beta+a_{0}-1=1$, this gives
$\mathbb{E}|X_{0}|\le\eps n/50+2$. For $1\le j\le r-2$, let $C$ be the constant of \Cref{lem:exposure},
which depends only on $d$ and $r$. With Case~2 and $T\log n=\eps n^{\beta}$, that lemma gives
$\mathbb{E}|X_{j}|\le C\log n+C\eps\,(b_{j}/b_{j-1})\,n^{\beta}(b_{j-1}/n)^{1/D}$.
Since $b_{j}\le n^{a_{j}}$, $b_{j-1}>n^{a_{j-1}}/2$, and $1-1/D>0$,
\[
  \frac{b_{j}}{b_{j-1}}\,n^{\beta}\Bigl(\frac{b_{j-1}}{n}\Bigr)^{1/D}
   =n^{\beta-1/D}\,b_{j}\,b_{j-1}^{-(1-1/D)}
   \le 2\,n^{\beta+a_{j}-a_{j-1}+(a_{j-1}-1)/D}=2n,
\]
where the last step is \Cref{eq:bal-ref}. Hence $\mathbb{E}|X_{j}|\le C\log n+2C\eps n$. By
\Cref{obs:badset-size} and $|\mathcal{A}_{-1}|<b_{0}\le n^{a_{0}}$,
$\mathbb{E}|\mathcal{A}|\le 3^{r-1}\bigl(n^{a_{0}}+(2C+1/50)\,\eps n+C\log n+2\bigr)$. Choose $\eps$ with
$3^{r-1}(2C+1/50)\,\eps\le 1/128$. Then $\mathbb{E}|\mathcal{A}|\le n/64$ for large $n$, since $a_{0}<1$.

\emph{Survivors.} Every position lies in fewer than $g$ cells of $\mathcal{C}_{g}$, so at most
$g|\mathcal{A}|$ cells of $\mathcal{C}_{g}$ have a bad position. Since
$\mathbb{E}[g|\mathcal{A}|]\le ng/64\le|\mathcal{C}_{g}|/8$, Markov's inequality applied to the
number of these cells shows that, with probability at least $1/2$, at most $|\mathcal{C}_{g}|/4$
cells of $\mathcal{C}_{g}$ have a bad position. By \Cref{lem:scatter} at scale $g$, with probability
at least $7/8$ at least half of the cells $w\in\mathcal{C}_{g}$ satisfy
$\dist(w,\mathcal{C}_{g}\setminus\{w\})>\rho$, where $\rho=\Theta_{d}\bigl((n/g)^{1/D}\bigr)$. On the
intersection of the two events, which has probability at least $1-\tfrac12-\tfrac18=\tfrac38$, let
$\mathcal{H}$ be the set of cells of $\mathcal{C}_{g}$ that have two clean positions and satisfy this
condition. Then $|\mathcal{H}|\ge|\mathcal{C}_{g}|/4\ge ng/32$, and any two cells of $\mathcal{H}$
are more than $\rho$ apart.

\emph{Forcing.} Let $\{u,v\}\in\mathcal{H}$. Both $u$ and $v$ are clean, so by \Cref{obs:clean}(b)
neither has been compared, in rounds $1,\dots,r-1$, with another position of its scale-$g$ block.
By \Cref{lem:prigid}, applied to $P_{r-1}$, the positions $u$ and $v$ are incomparable in $P_{r-1}$.
The revealed scales $b_{0},\dots,b_{r-3}$ are coarser than $g$, and $u$ and $v$ are co-blocked at
scale $g$, so they have the same label at every revealed scale. By \Cref{lem:pfinal}, the route of
round $r$ visits $\{u,v\}$.

\emph{Cost of the Final Round.} By the previous stage, the route of round $r$ visits every cell of
$\mathcal{H}$. Since $|\mathcal{H}|\ge 2$ and the cells of $\mathcal{H}$ are pairwise more than
$\rho$ apart, \Cref{obs:model}(b) and \Cref{fact:nn} give
\[
  L_{r}\ \ge\ \tfrac12|\mathcal{H}|\rho
   \ =\ \Omega_{d}\bigl(ng\,(n/g)^{1/D}\bigr)
   \ =\ \Omega_{d}\bigl(n^{1+a_{r-2}+(1-a_{r-2})/D}\bigr)
   \ =\ \Omega_{d}(n^{\beta}),
\]
where the second step uses $g>n^{a_{r-2}}/2$ and the last is \Cref{eq:bal-clean}. So
$L_{r}=\Omega_{d}(n^{\beta})$ on an event of probability at least $3/8$.

In Case~2 this gives $\mathbb{E}[L_{r}]=\Omega_{d}(n^{\beta})$, and in Case~1 we have
$\mathbb{E}[\mathrm{Move}]\ge T/100=\eps n^{\beta}/(100\log n)$. In both cases
$\mathbb{E}[\mathrm{Move}]=\Omega_{d,r}(n^{\beta}/\log n)$, which proves \Cref{thm:lb}.
\end{proof}

\begin{remark}
\label{rem:free}
\Cref{thm:lb} holds with free computation and memory: comparisons are the algorithm's only access to
the order. After Yao's principle fixes
a deterministic algorithm, the proof of \Cref{lem:pfinal} gives, for every pair of positions with
the same revealed labels that the first $r-1$ rounds leave incomparable, two inputs that agree on all
revealed labels and on all outcomes of these rounds but order the pair oppositely. Hence no computation on these outcomes orders the pair, and
the comparisons of the final round, with their movement, are required of every correct algorithm.
\end{remark}

\subsection{Exact Order on the Line}
\label{subsec:line-exact}

The logarithmic factor in \Cref{thm:lb} comes from \Cref{lem:density} alone, and on the line we remove it. In this
subsection $d=1$, and we write $\beta_{r}=\beta_{r}^{(1)}$.

\begin{theorem}
\label{thm:line-exact}
For $d=1$ and every fixed $r\ge 1$, $\Mof{r}{1}(n)=\Omega_{r}\bigl(n^{\beta_{r}}\bigr)$.
\end{theorem}

Together with \Cref{thm:line-ub}, it gives $\Mof{r}{1}(n)=\Theta_{r}\bigl(n^{\beta_{r}}\bigr)$.

The logarithm in \Cref{lem:density} comes from a product box that holds about $\log n$ cells, and such crowded
places are rare. So we group nearby cells into clusters. Small clusters are far apart for their size, so a route must
move to visit their cells, and large clusters almost never occur. The bound that replaces \Cref{lem:density} has an
additive term $\Xi_{b}$, which counts the cells of large clusters and which we define first.

Fix a dyadic scale $b\le\sqrt{n}$, and identify each cell with its fold, a point of
$\{(x,y):x<y\}$. By \Cref{obs:model}(b), the distance $\dist$ between two cells is the $\ell_{1}$ distance between their folds. Let $H=\lceil\sqrt{n/b}\rceil$,
let $\eps_{\star}>0$ and $n_{0}$ be absolute constants chosen in the proof of \Cref{lem:xi-bound},
and put
\[
  \rho_{t}=\eps_{\star}H\,4^{t}\quad(0\le t\le t_{\star}),\qquad
  t_{\star}=\lceil\log\log n\rceil,\qquad \kappa=2^{t_{\star}}.
\]
A \emph{cluster} is a set of cells of $\mathcal{C}_{b}$. Clusters are formed in phases
$t=0,1,\dots,t_{\star}$, and the clusters of phase $0$ are the single cells of $\mathcal{C}_{b}$.
Two clusters of phase $t$ are \emph{close} if some cell of one lies within distance $\rho_{t}$ of
some cell of the other. Phase $t$ merges each connected component of the closeness relation that
contains two or more clusters into one cluster of phase $t+1$. A cluster of phase $t$ that is close
to no other cluster is \emph{isolated}. An isolated cluster \emph{retires} if it has at most
$2^{t+1}$ cells. Otherwise it is also a cluster of phase $t+1$. The \emph{residual clusters} are the
clusters of phase $t_{\star}+1$. For $n\ge n_{0}$, let $\Xi_{b}$ be the total number of cells in residual
clusters, and for $n<n_{0}$, let $\Xi_{b}=0$.

By induction on $t$, every cluster of phase $t$ has at least $2^{t}$ cells: a merged component
consists of at least two clusters with at least $2^{t}$ cells each, and an isolated cluster that
does not retire has more than $2^{t+1}$ cells. In particular every residual cluster has at least
$2\kappa$ cells, and every cell of $\mathcal{C}_{b}$ ends in exactly one retired or residual cluster.

From one phase to the next, the distance $\rho_{t}$ grows by a factor of $4$, while the largest size
of a retiring cluster only doubles. So the clusters that retire in a phase are far apart for their
size.

\begin{lemma}
\label{lem:sharp-density}
Let $d=1$. There are absolute constants $C$ and $C'$ such that, for every input and every dyadic
scale $b\le\sqrt{n}$, every route of movement $L$ visits at most $C\,L\sqrt{b/n}+\Xi_{b}+C'\log n$
cells of $\mathcal{C}_{b}$.
\end{lemma}

\Cref{lem:sharp-density} is the analog of \Cref{lem:density} on the line, and the next lemma shows
that its additive term $\Xi_{b}$ is at most $1$ in expectation.

\begin{lemma}
\label{lem:xi-bound}
Let $d=1$. For a suitable choice of the absolute constants $\eps_{\star}>0$ and $n_{0}$,
$\mathbb{E}[\Xi_{b}]\le 1$ over the uniform input, for every dyadic scale $b\le\sqrt{n}$.
\end{lemma}

With these two lemmas, \Cref{lem:exposure} has the following analog on the line.

\begin{lemma}
\label{lem:exposure-line}
We have $\mathbb{E}|X_{0}|\le 2(L_{1}+1)\,b_{0}/n$, and for every
$1\le j\le r-2$,
\[
  \mathbb{E}|X_{j}|\le\frac{8b_{j}}{b_{j-1}}
    \Bigl(C\,\mathbb{E}[L_{j+1}]\Bigl(\frac{b_{j-1}}{n}\Bigr)^{1/2}+C'\log n+1\Bigr),
\]
where $C$ and $C'$ are the constants of \Cref{lem:sharp-density}.
\end{lemma}

\begin{proof}
The bound on $\mathbb{E}|X_{0}|$ is that of \Cref{lem:exposure}. Let $1\le j\le r-2$. At $q=1/2$ we
have $\delta=1/\bigl(2(1-2^{-r})\bigr)>1/2$, so $a_{0}=1-\delta<1/2$ and
$b_{j-1}\le b_{0}<\sqrt{n}$, so \Cref{lem:sharp-density} applies at scale $b_{j-1}$. We follow
the proof of \Cref{lem:exposure} with \Cref{lem:sharp-density} in place of \Cref{lem:density}. As
there, $\mathbb{E}\bigl[|X_{j}|\bigm|\mathcal{F}_{j}\bigr]\le(8b_{j}/b_{j-1})|K_{j}|$. By
\Cref{lem:sharp-density}, $|K_{j}|\le C\,L_{j+1}(b_{j-1}/n)^{1/2}+\Xi_{b_{j-1}}+C'\log n$ on every
input. Taking expectations and using $\mathbb{E}[\Xi_{b_{j-1}}]\le 1$ (\Cref{lem:xi-bound}) gives the
bound.
\end{proof}

\begin{proof}[Proof of \Cref{thm:line-exact}]
For $r\le 2$ the theorem is \Cref{prop:endpoints}, so let $r\ge 3$. As in the proof of
\Cref{thm:lb}, it suffices to show that the algorithm fixed in \Cref{subsec:badset} has
$\mathbb{E}[\mathrm{Move}]=\Omega_{r}(n^{\beta_{r}})$. We follow the proof of \Cref{thm:lb} with three
changes.

First, $T=\eps_{r}n^{\beta_{r}}$, where $\eps_{r}>0$ depends only on $r$ and is chosen below.
Cases~1 and~2 are defined with this $T$.

Second, the first stage uses \Cref{lem:exposure-line} in place of \Cref{lem:exposure}. In Case~2 it
gives $\mathbb{E}|X_{0}|\le 2(T/100+1)b_{0}/n\le\eps_{r}n/50+2$, as in the proof of \Cref{thm:lb},
and, for $1\le j\le r-2$,
\[
  \mathbb{E}|X_{j}|
   \le\frac{8b_{j}}{b_{j-1}}\Bigl(\frac{C\,T}{100}\Bigl(\frac{b_{j-1}}{n}\Bigr)^{1/2}+C'\log n+1\Bigr)
   \le\frac{C}{4}\,\eps_{r}n+8C'\log n+8,
\]
because $(b_{j}/b_{j-1})\,n^{\beta_{r}}(b_{j-1}/n)^{1/2}\le 2n$ by the computation in the proof of
\Cref{thm:lb} with $D=2$. By \Cref{obs:badset-size},
$\mathbb{E}|\mathcal{A}|\le 3^{r-1}\bigl(b_{0}+(C/4+1/50)\,\eps_{r}n+8C'\log n+10\bigr)$.

Third, choose $\eps_{r}$ with $3^{r-1}(C/4+1/50)\,\eps_{r}\le 1/128$. Then
$\mathbb{E}|\mathcal{A}|\le n/64$ for large $n$. The remaining stages use only this bound, and they
give $\mathbb{E}[L_{r}]=\Omega_{r}(n^{\beta_{r}})$ in Case~2. In Case~1,
$\mathbb{E}[\mathrm{Move}]\ge\eps_{r}n^{\beta_{r}}/100$. Hence
$\mathbb{E}[\mathrm{Move}]=\Omega_{r}(n^{\beta_{r}})$, which proves \Cref{thm:line-exact}.
\end{proof}

\begin{proof}[Proof of \Cref{lem:sharp-density}]
If $n<n_{0}$, then $\Xi_{b}=0$, and a route visits at most $|\mathcal{C}_{b}|\le nb/2\le n^{3/2}\le
n_{0}^{3/2}$ cells of $\mathcal{C}_{b}$. Let $n\ge n_{0}$. Deleting the configurations $(x,x)$ from
a route of movement $L$ and folding the others gives a sequence of folds of $\ell_{1}$-length at most
$L$ (\Cref{obs:model}(b)), and the cells of $\mathcal{C}_{b}$ that the route visits are those whose folds occur in this sequence. The
residual clusters contain $\Xi_{b}$ cells in all, and every other cell of $\mathcal{C}_{b}$ lies in
a retired cluster. Fix a phase $t$. Two clusters that retire in phase $t$ are isolated in phase
$t$, so every cell of one is more than $\rho_{t}$ from every cell of the other. Order the clusters
retired in phase $t$ that have a cell in the sequence by the first occurrence of one of their cells.
Consecutive first occurrences are more than $\rho_{t}$ apart, so at most $1+L/\rho_{t}$ of these
clusters have a cell in the sequence, and each of them has at most $2^{t+1}$ cells. Summing over the
phases, the route visits at most
\[
  \sum_{t=0}^{t_{\star}}2^{t+1}\Bigl(1+\frac{L}{\rho_{t}}\Bigr)
   \le 2^{t_{\star}+2}+\frac{L}{\rho_{0}}\sum_{t\ge 0}\frac{2^{t+1}}{4^{t}}
   =2^{t_{\star}+2}+\frac{4L}{\rho_{0}}
\]
cells of retired clusters. Since $2^{t_{\star}}<2\log n$ and $\rho_{0}=\eps_{\star}H\ge
\eps_{\star}\sqrt{n/b}$, this is at most $8\log n+(4/\eps_{\star})L\sqrt{b/n}$. Adding the $\Xi_{b}$
cells of residual clusters gives the bound for $n\ge n_{0}$. Hence the lemma holds with
$C=4/\eps_{\star}$ and $C'=\max\{8,n_{0}^{3/2}\}$.
\end{proof}

\begin{proof}[Proof of \Cref{lem:xi-bound}]
If $n<n_{0}$, then $\Xi_{b}=0$, so let $n\ge n_{0}$, where $n_{0}$ is chosen at the end of the
proof. We charge $\Xi_{b}$ to sets of between $\kappa$ and $2\kappa$ cells of $\mathcal{C}_{b}$ that
carry a tree of short edges, bound the probability that a fixed set of cells lies in
$\mathcal{C}_{b}$, and bound the expected number of such sets.

For a finite set $W$ of cells, its \emph{position graph} has the positions occurring in $W$ as
vertices and the cells of $W$ as edges. Let $c(W)$ be its number of vertices minus its number of
components, which is the number of rank constraints that $W\subseteq\mathcal{C}_{b}$ imposes, and let $h(W)=|W|-c(W)$ be its cyclomatic number. We say that $W$ is \emph{eligible} if
some tree with vertex set $W$ carries a phase $\ell(e)\in\{0,\dots,t_{\star}\}$ on each of its
edges $e=\{w,w'\}$ such that $\lVert w-w'\rVert_{1}\le\rho_{\ell(e)}$ for every edge and
$\sum_{e}\ell(e)\le C_{0}|W|$, where $C_{0}=16$. A \emph{witness} is an eligible subset of
$\mathcal{C}_{b}$ with between $\kappa$ and $2\kappa$ cells. Let $N_{s}$ be the number of eligible subsets of
$\mathcal{C}_{b}$ with $s$ cells.

\emph{Charging.} Let $F$ be a residual cluster. Its merge history gives a tree with vertex set $F$:
when phase $t$ merges a component of close clusters, choose a spanning tree of the component and,
for each of its edges, a pair of cells, one in each of the two clusters, at distance at most
$\rho_{t}$, and these pairs become tree edges with phase $t$. A merge in phase $t$ adds fewer edges than
it merges clusters, and every cluster of phase $t$ has at least $2^{t}$ cells, so at most
$2^{-t}|F|$ tree edges have phase $t$, and the phases sum to at most
$\sum_{t\ge 0}t\,2^{-t}|F|=2|F|$.

We cut this tree into connected pieces. Root the tree and collect edges from
the leaves upward: at each vertex, add the edge to each child together with the edges left over at
that child, and emit the collected edges as one piece as soon as they number at least $\kappa-1$.
Every collected set is connected, and the edges left over at a vertex number fewer than $\kappa-1$,
so each piece has between $\kappa-1$ and $2\kappa-3$ edges and hence between $\kappa$ and $2\kappa$
cells. The pieces have disjoint edge sets, and fewer than $\kappa-1$ edges remain at the root. The
tree has $|F|-1$ edges and $|F|\ge 2\kappa$, so there are more than
$(|F|-\kappa)/(2\kappa)\ge|F|/(4\kappa)$ pieces.
Their phase sums total at most $2|F|$, so fewer than $|F|/(8\kappa)$ pieces have phase sum above
$16\kappa$. Each of the remaining, more than $|F|/(8\kappa)$, pieces has a cell set $W$ with
$|W|\ge\kappa$ and phase sum at most $16\kappa\le C_{0}|W|$, so $W$ is a witness. Distinct pieces
have distinct cell sets, since a connected subtree with cell set $W$ contains every tree edge inside
$W$, and distinct residual clusters are disjoint. Hence each residual cluster $F$ yields more than
$|F|/(8\kappa)$ witnesses, all distinct, and
\[
  \Xi_{b}\ \le\ 8\kappa\sum_{s=\kappa}^{2\kappa}N_{s}.
\]

\emph{Witness Probability.} Fix a set $W$ of $s\le 2\kappa$ cells. Reveal the ranks of its positions
one component of the position graph at a time, each component in an order in which every vertex $x$
after the first has a neighbor $x'$ earlier in the order. If $W\subseteq\mathcal{C}_{b}$, then every
such $x$ receives a rank in the scale-$b$ block of $x'$. Given the earlier ranks, this block contains
at most $b$ of the at least $n-2s\ge n/4$ unrevealed ranks, so this happens with probability at most $4b/n$. There are $c(W)$ such vertices, so
$\Pr[W\subseteq\mathcal{C}_{b}]\le(4b/n)^{c(W)}=(4b/n)^{s-h(W)}$.

\emph{First Moment.} Let $N_{s}^{(h)}$ be the number of eligible subsets of $\mathcal{C}_{b}$ with
$s$ cells and cyclomatic number $h$. We bound $\mathbb{E}[N_{s}^{(h)}]$ for $h\le\eta s/t_{\star}$
here, where $\eta$ is chosen below, and treat larger $h$ in the next paragraph. Given an eligible set
$W$ with $s$ cells, fix a tree as in the definition of eligibility, root it, order the children of
each vertex, and add the cells of $W$ one at a time in the preorder of this ordered rooted tree, so
that every cell comes after its parent. Adding a cell increases $c$ of the set added so far by one,
unless its two positions already lie in one component of the position graph of that set. We then say
that the cell \emph{closes a cycle}. Since $c(W)=s-h$, exactly $h$ cells close a cycle. Hence $W$ is encoded by the ordered rooted tree,
at most $4^{s}$ choices; the phases of its edges, which sum to at most $C_{0}s$, at most
$2^{(C_{0}+1)s}$ choices; the set of cells that close a cycle, at most $2^{s}$ choices; and the
placement of the cells in preorder. A cell that closes a cycle is one of at most $(2s)^{2}$ pairs of
positions already present. The root is one of at most $n^{2}/2$ cells. Every other cell lies within
distance $\rho_{\ell}$ of its parent, where $\ell$ is the phase of the tree edge to its parent, so it
has at most $(2\rho_{\ell}+1)^{2}\le 9\rho_{\ell}^{2}$ placements, since
$\rho_{\ell}\ge\eps_{\star}H\ge\eps_{\star}n^{1/4}\ge 1$ for $n\ge n_{0}$. The bound
$\Pr[W\subseteq\mathcal{C}_{b}]\le(4b/n)^{s-h}$ gives a factor $4b/n$ for each of the $s-h$ cells
that close no cycle. Since $H^{2}\le 4n/b$, we have
$9\rho_{\ell}^{2}\cdot 4b/n\le 144\,\eps_{\star}^{2}16^{\ell}$, and the phases give
$\prod_{e}16^{\ell(e)}\le 16^{C_{0}s}$. Collecting the factors,
\[
  \mathbb{E}\bigl[N_{s}^{(h)}\bigr]
   \le 2nb\;\eps_{\star}^{-2}\,\bigl(C_{1}\eps_{\star}^{2}\bigr)^{s}\,
       \bigl(4s^{2}/\eps_{\star}^{2}\bigr)^{h}
\]
for an absolute constant $C_{1}$. Let $\eps_{\star}^{2}=1/(16C_{1})$ and
$\eta=\bigl(6+2\log(1/\eps_{\star})\bigr)^{-1}$. If $h\le\eta s/t_{\star}$, then
$(4s^{2}/\eps_{\star}^{2})^{h}\le 2^{s}$. Indeed, $s\le 2\kappa<4\log n$ gives
$\log(4s^{2})\le 6+2\log\log n\le 6t_{\star}$ for $n\ge n_{0}$, so
$h\log(4s^{2}/\eps_{\star}^{2})\le(\eta s/t_{\star})\bigl(6t_{\star}+2\log(1/\eps_{\star})\bigr)\le s$.
Hence $\mathbb{E}[N_{s}^{(h)}]\le 2nb\,\eps_{\star}^{-2}8^{-s}$, and summing over the at most $s$
values of $h$ gives $\sum_{h\le\eta s/t_{\star}}\mathbb{E}[N_{s}^{(h)}]\le nb\,4^{-s}$ for
$s\ge\kappa$ and $n\ge n_{0}$.

\emph{High Cyclomatic Number.} Let $N_{s}^{\mathrm{high}}$ be the number of eligible subsets $W$ of
$\mathcal{C}_{b}$ with $s$ cells and $h(W)>\eta s/t_{\star}$. For these we use the geometry of the
set instead. Tree edges join cells at distance at most $\rho_{t_{\star}}=\eps_{\star}H\kappa^{2}$, so
every cell of $W$ lies within distance $(s-1)\rho_{t_{\star}}$ of the root, and all positions of $W$
lie in the two intervals of radius $R=s\rho_{t_{\star}}$ around the two positions of the root. These
intervals contain $V\le 4R+2=O\bigl(\sqrt{n/b}\,\log^{3}n\bigr)$ positions. Let $N_{s,\omega,\tau}$ be
the number of these sets whose position graph has $\omega$ vertices and $\tau$ components. Such a set
is encoded by its root, at most $n^{2}$ choices, its other $\omega-2$ positions, at most
$V^{\omega-2}$ choices, and its $s$ cells as pairs of its positions, at most $(2s)^{2s}$ choices.
With $\gamma=4b/n$ and $z=\omega-2\tau$, the bound on $\Pr[W\subseteq\mathcal{C}_{b}]$ gives
\[
  \mathbb{E}[N_{s,\omega,\tau}]\le n^{2}(2s)^{2s}V^{\omega-2}\gamma^{\omega-\tau}
   =n^{2}(2s)^{2s}V^{-2}\,\bigl(V^{2}\gamma\bigr)^{\tau}\,\bigl(V\gamma\bigr)^{z}.
\]
Here $V^{2}\gamma=O(\log^{6}n)$, and $V\gamma\le n^{-1/5}$ for $n\ge n_{0}$ because $b\le\sqrt{n}$.
The position graph is simple, so a component with two vertices contributes nothing to $z$ or to $h$.
A component with $\omega'\ge 3$ vertices contributes $\omega'-2\ge\omega'/3$ to $z$, so
$z\ge\omega_{3}/3$, where $\omega_{3}$ is the number of vertices in components with at least three vertices. Also
$h\le\omega_{3}^{2}/2$, because $h$ is at most the number of edges in these components. Hence
$z\ge\sqrt{2h}/3$, and $h>\eta s/t_{\star}\ge\eta\kappa/t_{\star}$ gives
$z=\Omega\bigl(\sqrt{\log n/\log\log n}\bigr)$. So $(V\gamma)^{z}=n^{-\Omega(\sqrt{\log n/\log\log n})}$,
whereas $n^{2}(2s)^{2s}(V^{2}\gamma)^{\tau}=\exp\bigl(O(\log n\,\log\log n)\bigr)$ because
$s,\tau\le 4\log n$. For $n\ge n_{0}$ the product is at most $n^{-22}$, and summing over the at most
$(2s)^{2}\le n^{2}$ pairs $(\omega,\tau)$ gives $\mathbb{E}[N_{s}^{\mathrm{high}}]\le n^{-20}$.

For $\kappa\le s\le 2\kappa$ the two regimes give $\mathbb{E}[N_{s}]\le nb\,4^{-s}+n^{-20}$. Since
$4^{-\kappa}\le n^{-2}$ and $b\le\sqrt{n}$,
\[
  \mathbb{E}[\Xi_{b}]\ \le\ 8\kappa\sum_{s=\kappa}^{2\kappa}\mathbb{E}[N_{s}]
   \ \le\ 8\kappa(\kappa+1)\bigl(n^{-1/2}+n^{-20}\bigr)\ \le\ 1,
\]
where $n_{0}$ is chosen so large that this inequality and the estimates above hold for all
$n\ge n_{0}$.
\end{proof}

\section{Routing and the Sampling Algorithm}
\label{sec:cascade}

This section and the next prove the upper-bound half of \Cref{thm:main}. Every route of our algorithms
visits configurations known when the round starts, and \Cref{subsec:routing} bounds the movement this takes. A
merge sort then gives the fully adaptive bound (\Cref{subsec:floor}). For a fixed number $r\ge 2$ of rounds we use the
sampling algorithm, which refines the rank order round by round (\Cref{subsec:cascade}). On the line, each refinement round
selects intervals of one length (\Cref{subsec:line}).

\subsection{Routing}
\label{subsec:routing}

The next lemma bounds the movement needed to visit a known set of configurations.

\begin{lemma}
\label{lem:batch}
\begin{enumerate}[label=\textup{(\roman*)}]
  \item Let $\mathcal{Q}$ be a set of $Q\ge 1$ points of a box in $\mathbb{Z}^{D'}$ whose sides are at
        most $m'$. From any given point of the box, a lattice path inside the box, that is, a
        sequence of points with consecutive points at $\ell_{1}$ distance $1$, visits all points of
        $\mathcal{Q}$ and has length $O_{D'}\bigl(m'Q^{1-1/D'}+m'\bigr)$. The path can be computed
        from $\mathcal{Q}$ in polynomial time.
  \item In the comparison grid, any $Q\ge 1$ configurations are visited by a route that starts at
        the current configuration and has movement $O_{d}\bigl(n^{1/d}Q^{1-1/D}+n^{1/d}\bigr)$.
\end{enumerate}
\end{lemma}

\begin{proof}
For~(i), we cut the box into small cubes and traverse them in serpentine order, visiting the points
of $\mathcal{Q}$ in each cube on the way. Let $\Delta=\lceil m'Q^{-1/D'}\rceil$ and cut the box into
cubes of side $\Delta$. There are at most $(m'/\Delta+1)^{D'}\le(Q^{1/D'}+1)^{D'}\le 2^{D'}Q$ of
them. Consecutive cubes of the serpentine order are adjacent, so the path moves $O_{D'}(\Delta)$
from one cube to the next. Inside a cube it moves $O_{D'}(\Delta)$ from one point of $\mathcal{Q}$
to the next. Together with the $O_{D'}(m')$ steps from the given point to the first cube, this gives
length $O_{D'}(\Delta Q+m')$. Since the box has at most $m'^{D'}$ points, $Q^{1/D'}\le m'$, so
$\Delta Q\le m'Q^{1-1/D'}+Q\le 2m'Q^{1-1/D'}$.

Part~(ii) is~(i) for the box $[m]^{D}$, with $D'=D$ and $m'=m=n^{1/d}$.
\end{proof}

Merging two sorted lists by walking them in rank order can cost a long jump per comparison. The next
lemma instead advances many binary searches in one round, so that one route serves them all.

\begin{lemma}
\label{lem:merge}
Let $R\subseteq[m]^{d}$ be a box of aspect ratio $O_{d}(1)$ with $|R|\ge 2$, and let its positions be
partitioned into two lists, each sorted by rank. A deterministic algorithm finds the rank order of
$R$ in $O\bigl((\log|R|)^{2}\bigr)$ rounds with total movement $O_{d}\bigl(|R|^{1+1/(2d)}\bigr)$.
Its routes stay in $R\times R$, and the bound holds whichever configuration of $R\times R$ each round
starts from.
\end{lemma}

\begin{proof}
The algorithm splits the two lists at the median of their union by binary search, recurses on the
two parts, and advances all binary searches of one depth together. Write $N=|R|$. A subproblem consists of a sublist of consecutive entries of each
list, $s$ entries in all, and asks for the $t=\lceil s/2\rceil$ lowest keys of their union. For
$1\le i\le t$, the $i$-th entry of the first sublist, if it exists, has $i-1$ entries of its sublist below it, so it
is among these $t$ keys exactly when at most $t-i$ entries of the second sublist lie below it.
Equivalently, it lies below the $(t-i+1)$-th entry of the second sublist, or that entry does not
exist. This condition is monotone in $i$, so a binary search on $i$ finds the $t$ lowest keys with
$O(\log(s+2))$ comparisons of one entry of each sublist. The $t$ lowest keys form a prefix of each sublist, so the two parts are
again subproblems, of sizes $\lceil s/2\rceil$ and $\lfloor s/2\rfloor$. The recursion stops at subproblems with an empty
sublist, which are already sorted, and this gives the rank order of $R$.

Each round advances every binary search of the current depth by one comparison. At depth $\lambda$
there are at most $2^{\lambda}$ subproblems of size at most $\lceil N/2^{\lambda}\rceil$, so this
depth takes $O\bigl(\log(N/2^{\lambda}+2)\bigr)$ rounds. There are $O(\log N)$ depths, which gives
the bound on the rounds. A round at depth $\lambda$ visits at most $2^{\lambda}$ known configurations
of $R\times R$, a box whose sides are $O_{d}(N^{1/d})$. By \Cref{lem:batch}(i), from any given
configuration of $R\times R$ its movement is $O_{d}\bigl(N^{1/d}2^{(1-1/D)\lambda}+N^{1/d}\bigr)$.
Substituting $i=\lceil\log_{2}N\rceil-\lambda$, the main terms total
\[
  O_{d}\Bigl(N^{1/d}\sum_{\lambda}2^{(1-1/D)\lambda}\log\bigl(N/2^{\lambda}+2\bigr)\Bigr)
   =O_{d}\Bigl(N^{1/d}N^{1-1/D}\sum_{i\ge 0}(i+2)\,2^{-(1-1/D)i}\Bigr)
   =O_{d}\bigl(N^{1+1/(2d)}\bigr),
\]
since the series converges and $1/d-1/D=1/(2d)$. The additive terms total
$O_{d}\bigl(N^{1/d}(\log N)^{2}\bigr)=O_{d}\bigl(N^{1+1/(2d)}\bigr)$.
\end{proof}

\subsection{The Fully Adaptive Bound}
\label{subsec:floor}

For $r=\infty$, a merge sort over a partition of the grid into boxes gives the upper bound in every
dimension. Together with \Cref{prop:floor}, it determines $\Mof{\infty}{d}(n)$.

\begin{proposition}
\label{prop:floor-ub}
For every $d\ge 1$, a deterministic algorithm sorts in $O\bigl((\log n)^{3}\bigr)$ rounds with movement
$O_{d}\bigl(n^{1+1/(2d)}\bigr)$. Hence $\Mof{\infty}{d}(n)=O_{d}\bigl(n^{1+1/(2d)}\bigr)$.
\end{proposition}

\begin{proof}
We sort by merging along a tree of boxes of depth $T=d\lceil\log_{2}m\rceil\le\log_{2}n+d$. The root
is $[m]^{d}$. At depth $t$, every box whose side along axis $t\bmod d$ is $s\ge 2$ is halved along
that axis, into sides $\lceil s/2\rceil$ and $\lfloor s/2\rfloor$. A box whose side along this axis is
$1$ has itself as its only child. At depth $T$ all boxes are single positions. The boxes of one depth
form a grid and have aspect ratio $O(1)$. At height $i$, that is, at depth $T-i$, there are at most
$2^{T-i}\le 2^{d}n/2^{i}$ boxes, and each has at most $2^{i}$ positions, because a box has at most two
children. The algorithm processes the heights in increasing order and merges the two halves of each
halved box by \Cref{lem:merge}.

By \Cref{lem:merge}, the movement inside the boxes of height $i$ is
$O_{d}\bigl((n/2^{i})\cdot 2^{i(1+1/(2d))}\bigr)=O_{d}\bigl(n\,2^{i/(2d)}\bigr)$. Summing over
$i\le T$ gives a geometric series dominated by its last term, so the total is
$O_{d}\bigl(n\,2^{T/(2d)}\bigr)=O_{d}\bigl(n^{1+1/(2d)}\bigr)$.

The merges of one height share their rounds: a round of the algorithm is one round of each merge, and
its route visits the boxes in the serpentine order of their grid. The heads cross from one box to
the next with movement $O_{d}(2^{i/d})$, so these crossings, together with the move to the first box,
add $O_{d}\bigl((n/2^{i})\,2^{i/d}+n^{1/d}\bigr)=O_{d}(n)$ per round. Each merge takes
$O\bigl((\log n)^{2}\bigr)$ rounds and there are $O(\log n)$ heights, so the algorithm has $O\bigl((\log n)^{3}\bigr)$ rounds
and the crossings total $O_{d}\bigl(n(\log n)^{3}\bigr)=O_{d}\bigl(n^{1+1/(2d)}\bigr)$.
\end{proof}

\begin{remark}
\label{rem:floor-context}
At $d=1$ the tree of boxes is the merge tree of ordinary merge sort, with the merges of \Cref{lem:merge}. The exponent $3/2$ agrees with the bound of Maass and
Schnitger for sorting on a Turing machine with one worktape and a two-way read-only input
tape~\cite{MS86,DMS91}. Their input consists of bit strings, so the agreement is up to logarithmic
factors.
\end{remark}

\subsection{The Sampling Algorithm}
\label{subsec:cascade}

Between rounds, the sampling algorithm knows the
rank order only up to a partition of the keys into sets of consecutive ranks.

\begin{definition}[Segment]
\label{def:segment}
A \emph{segment} is a nonempty set of positions whose ranks form an interval of $[n]$ known to the
algorithm. A segment may be a single position.
\end{definition}

The comparisons that remain are the pairs within segments, and the movement they take depends on
where each segment lies: a segment with many keys in a small region is cheap. The following
quantity records this, by weighting each key with the number of boxes that its segment meets.

\begin{definition}[Dispersion]
\label{def:dispersion}
For a partition $\mathcal{P}$ of $[m]^{d}$ into boxes and a family of disjoint segments, let
$\mathcal{P}(G)$ be the set of boxes of $\mathcal{P}$ that a segment $G$ meets. The \emph{dispersion}
of the family is
\[
  S_{\mathcal{P}}=\sum_{G}|G|\,|\mathcal{P}(G)|.
\]
\end{definition}

Splitting a segment never raises the dispersion: if $G_{1},G_{2}\subseteq G$ are disjoint, then
$\mathcal{P}(G_{i})\subseteq\mathcal{P}(G)$, so
$|G_{1}|\,|\mathcal{P}(G_{1})|+|G_{2}|\,|\mathcal{P}(G_{2})|\le|G|\,|\mathcal{P}(G)|$.

The algorithm has $r$ rounds. The first chooses random positions, the \emph{splitters}. The next
$r-2$ are \emph{refinement rounds}, which draw samples, and the last is the \emph{cleanup}. A
\emph{sample} is given by a partition $\mathcal{P}$ of $[m]^{d}$ into boxes and a rate $p\in(0,1]$. Drawing it selects each box of $\mathcal{P}$ independently with
probability $p$, and the keys in selected boxes that lie in segments are its \emph{pivots}. The
splitters and the pivots are called \emph{resolved keys}, since the order of each of them relative
to every other key is known after its round. Resolved keys lie in no segment.

\begin{algorithm}[t]
\caption{The sampling algorithm}
\label{alg:sampling}
\begin{algorithmic}[1]
\REQUIRE $r\ge 2$; the exponents $a_{0}>\dots>a_{r-2}$ of \Cref{lem:schedule}, with $b_{j}=n^{a_{j}}$;
  the samples of each refinement round; a route procedure for each refinement round and for the
  cleanup round.
\STATE \textbf{Round $1$ (splitter).} Choose $\lceil n/b_{0}\rceil$ positions uniformly at random
  without replacement, the splitters. \label{ln:split-draw}
\STATE The required configurations are all $(u,v)$ with $u\ne v$ and $v$ a splitter. For each
  splitter $v$ in turn, place the second head on $v$ and sweep the first head through $[m]^{d}$.
  \label{ln:split-route}
\STATE The splitters become resolved keys. In rank order, they cut the other positions into sets,
  and the nonempty sets are the segments. \label{ln:split-state}
\FOR{$j=1,\dots,r-2$}
  \STATE \textbf{Round $j+1$ (refinement).} Draw the samples of this round independently. The pivots
    of the round are the pivots of its samples. \label{ln:ref-draw}
  \STATE The required configurations are all $(u,v)$ with $u\ne v$ and $v$ a pivot in the
    segment of $u$. Visit them by the route procedure. \label{ln:ref-route}
  \STATE The pivots become resolved keys. In each segment, the nonempty maximal pivot-free sets of
    consecutive ranks become the new segments. \label{ln:ref-state}
\ENDFOR
\STATE \textbf{Round $r$ (cleanup).} The required configurations are, for each cell whose two positions lie
  in one segment, one fixed orientation. Visit them by the route procedure. \label{ln:clean-route}
\STATE Sort each segment, and output the resolved keys and the segments in rank order.
  \label{ln:output}
\end{algorithmic}
\end{algorithm}

\Cref{alg:sampling} is a template: each instance below specifies the samples and the routes. The
state between rounds is the sequence of resolved keys and nonempty segments in rank order, and the
analysis tracks the dispersion of the segments. The required configurations of a round are known when the round
starts, so its route depends only on the coins and on the outcomes of earlier rounds, and in every
instance it is computed in polynomial time. Line~\ref{ln:split-route} moves $O(n)$ per splitter, so
round $1$ has movement $O(n\lceil n/b_{0}\rceil)=O(n^{2-a_{0}})$, and $2-a_{0}=\beta_{r}^{(d)}$ by
\Cref{eq:bal-split}.

The algorithm is correct for every input and every outcome of its coins. Every key other than a
splitter is compared with all splitters in round $1$, so its segment is known. In a refinement round,
every key of a segment, pivots included, is compared with all other pivots of this segment, so the
new state is known. The cleanup compares all pairs within a segment, and the state determines the
order of every other pair.

The next lemma bounds the \emph{squared mass} $A=\sum_{G}|G|^{2}$ of the family after round $1$, which counts the pairs of
keys in a common segment. The
squared mass is the dispersion for the partition into single positions, where $|\mathcal{P}(G)|=|G|$.

\begin{lemma}
\label{lem:splitter}
Fix an input and choose $s$ positions uniformly at random without replacement, where $1\le s\le n$. In rank order, the chosen positions cut the other positions into sets $G_{0},\dots,G_{s}$, which may be empty. Then
$\mathbb{E}\sum_{i}|G_{i}|^{2}\le 3n^{2}/s$.
\end{lemma}

\begin{proof}
The sum $\sum_{i}|G_{i}|^{2}$ counts the ordered pairs of ranks that lie in a common set, and we bound
the probability of each pair. Since the input is a bijection, the chosen ranks form a uniformly random
set of $s$ ranks. Two ranks at distance $\Delta$ lie in a common set only if none of the $\Delta$ ranks starting at the lower one is
chosen, which has probability at most $(1-s/n)^{\Delta}\le\exp(-s\Delta/n)$. Hence
\[
  \mathbb{E}\sum_{i}|G_{i}|^{2}\ \le\ n+2n\sum_{\Delta\ge 1}\exp(-s\Delta/n)
   \ \le\ n+\frac{2n^{2}}{s}\ \le\ \frac{3n^{2}}{s},
\]
using $\sum_{\Delta\ge 1}\exp(-s\Delta/n)=(\exp(s/n)-1)^{-1}\le n/s$.
\end{proof}

A sample of rate $p$ leaves dispersion about $n/p$ for its own partition, in every dimension and whatever the
family that enters the round.

\begin{lemma}
\label{lem:sample}
Fix an input and a family of disjoint segments, and draw a sample with partition $\mathcal{P}$ and
rate $p$. Within each segment, the pivots split the other keys into \emph{refined segments}, the
nonempty maximal pivot-free sets of consecutive ranks. For a key $u$ in a segment, let $G_{u}$ be its
refined segment if $u$ is not a pivot, and $G_{u}=\emptyset$ otherwise. Then
$\mathbb{E}|\mathcal{P}(G_{u})|\le 3/p$, and hence the dispersion of the refined segments satisfies
$\mathbb{E}[S_{\mathcal{P}}]\le 3n/p$.
\end{lemma}

\begin{proof}
We bound the number of boxes that $G_{u}$ meets above and below $u$, and then sum over the keys. List
the segment of $u$ in rank order. Let $R_{u}^{+}$ be the number of boxes of $\mathcal{P}$, other
than the box of $u$, that the keys above $u$ meet before the first pivot above $u$. If
$R_{u}^{+}\ge t$, then the first $t$ such boxes in rank order are all unselected. These boxes are determined by the input and the family, and each is selected independently with probability $p$, so $\Pr[R_{u}^{+}\ge t]\le(1-p)^{t}$, and $\mathbb{E}[R_{u}^{+}]\le 1/p$.
Define $R_{u}^{-}$ in the same way for the keys below $u$. Then
$|\mathcal{P}(G_{u})|\le 1+R_{u}^{-}+R_{u}^{+}$, whose expectation is at most $3/p$. Finally,
$S_{\mathcal{P}}=\sum_{u}|\mathcal{P}(G_{u})|$, and there are at most $n$ keys.
\end{proof}

The later rounds tour the product boxes of a partition and visit the required configurations inside each of them. Fix a family of disjoint
segments and a partition $\mathcal{P}$ of $[m]^{d}$ into boxes. For $I,J\in\mathcal{P}$, the set
$I\times J$ is a product box, and we write $Q_{I,J}=\sum_{G}|G\cap I|\,|G\cap J|$. The product box
$I\times J$ is \emph{occupied} if $Q_{I,J}>0$, and $P$ is the number of occupied product boxes. If this
family enters a refinement round or the cleanup, a required configuration in $I\times J$ is a pair
$(u,v)$ with $u\in I$, $v\in J$, and $u$ and $v$ in a common segment, so at most $Q_{I,J}$ required
configurations lie in $I\times J$. The number of occupied product boxes bounds the tour, and on the line the sum of the
$\sqrt{Q_{I,J}}$ bounds the visits. The dispersion bounds both:
\begin{equation}
  \label{eq:macro}
  P\ \le\ \sum_{G}|\mathcal{P}(G)|^{2}\ \le\ S_{\mathcal{P}},
  \qquad
  \sum_{I,J}\sqrt{Q_{I,J}}\ \le\ \sum_{G}\Bigl(\sum_{I}\sqrt{|G\cap I|}\Bigr)^{2}\ \le\ S_{\mathcal{P}}.
\end{equation}
For the first chain, a segment contributes to $Q_{I,J}$ only if it meets both $I$ and $J$, which
holds for $|\mathcal{P}(G)|^{2}$ product boxes, and $|\mathcal{P}(G)|\le|G|$. For the second chain,
the first step is the subadditivity of the square root over the segments, and the second is the
Cauchy--Schwarz inequality over the $|\mathcal{P}(G)|$ nonzero terms $\sqrt{|G\cap I|}$.

\subsection{The Upper Bound on the Line}
\label{subsec:line}

On the line, the dispersion bounds both the tour of the next round and its visits (\Cref{eq:macro}), so each refinement round draws one sample. In this
subsection $d=1$, so $D=2$ and $q=\tfrac12$, and we write $\beta=\beta_{r}$. An
\emph{interval of length $h$} is a set of $h$ consecutive positions. We partition the line into
intervals of length $h$, the last possibly shorter, and write $\mathcal{P}_{h}(G)$ and $S_{h}$ for
this partition. An interval of length $h$ meets at most two intervals of length $h'\ge h$, so
$S_{h'}\le 2S_{h}$.

For $1\le j\le r-1$, let
$h_{j}=\lceil\sqrt{n/b_{j-1}}\rceil$, so that $h_{1}\le\dots\le h_{r-1}=\lceil\sqrt{n/g}\rceil$, and
for $1\le j\le r-2$, let $p_{j}=n^{-a_{j}}$. \Cref{alg:line} is the instance of
\Cref{alg:sampling} with these parameters. We also apply \Cref{lem:batch}(i) to coarser grids whose points are intervals, cubes, or product
boxes. A step of the lattice path then moves the heads to an adjacent interval, cube, or product box.

\begin{algorithm}[t]
\caption{The line algorithm}
\label{alg:line}
\begin{algorithmic}[1]
\REQUIRE $d=1$; $r\ge 2$; $h_{j}$ and $p_{j}$ as above.
\STATE \textbf{Round $1$ (splitter).} Lines~\ref{ln:split-draw}--\ref{ln:split-state} of
  \Cref{alg:sampling}.
\FOR{$j=1,\dots,r-2$}
  \STATE \textbf{Round $j+1$ (refinement).} Draw one sample: intervals of length $h_{j}$, rate
    $p_{j}$.
  \FOR{each selected interval $J$, in order on the line} \label{ln:line-ref-begin}
    \STATE Place the second head at the left endpoint of $J$, and sweep the first head along the
      line from left to right.
    \STATE When the first head reaches an interval $I$ with $Q_{I,J}>0$, visit the required configurations
      in $I\times J$ by \Cref{lem:batch}(i). Then return the second head to the left endpoint of
      $J$ and the first head to its place in the sweep. \label{ln:line-ref-end}
  \ENDFOR
  \STATE Update the state as in Line~\ref{ln:ref-state} of \Cref{alg:sampling}.
\ENDFOR
\STATE \textbf{Round $r$ (cleanup).} With $h=h_{r-1}$, move along the lattice path of
  \Cref{lem:batch}(i) through the occupied product boxes on the grid of $\lceil n/h\rceil^{2}$
  product boxes. On first arrival at an occupied $I\times J$, visit the required configurations in it by
  \Cref{lem:batch}(i). \label{ln:line-clean}
\STATE Output as in Line~\ref{ln:output} of \Cref{alg:sampling}.
\end{algorithmic}
\end{algorithm}

The next two lemmas bound the movement of these routes for any family of segments.

\begin{lemma}
\label{lem:refine}
Let a family of disjoint segments have dispersion $S_{h}$ for intervals of length $h$, and let
$p\in(0,1]$. For a sample with intervals of length $h$ and rate $p$, the route of
Lines~\ref{ln:line-ref-begin}--\ref{ln:line-ref-end} of \Cref{alg:line} visits all required configurations and
has expected movement $O\bigl(p\,n^{2}/h+p\,h\,S_{h}+n\bigr)$.
\end{lemma}

\begin{proof}
The route visits all required configurations, because every required configuration lies in a product box $I\times J$
with $J$ selected and $Q_{I,J}>0$. The first head moves at most $2n$ per selected interval on its
sweep, and the second head moves at most $2n$ between selected intervals over the whole round. In
expectation there are at most $p(n/h+1)$ selected intervals, so these moves total
$O(p\,n^{2}/h+n)$. For an occupied $I\times J$ with $J$ selected, at most $Q_{I,J}$ required configurations lie
in $I\times J$, so by \Cref{lem:batch}(i) visiting them and returning takes movement
$O\bigl(h\sqrt{Q_{I,J}}\bigr)$, because $Q_{I,J}\ge 1$. Each $J$ is selected with probability $p$, so by
\Cref{eq:macro} these visits have expected total $O\bigl(p\,h\sum_{I,J}\sqrt{Q_{I,J}}\bigr)=O(p\,h\,S_{h})$.
\end{proof}

\begin{lemma}
\label{lem:cleanup}
Let a family of disjoint segments have dispersion $S_{h}$ for intervals of length $h$. The route of
Line~\ref{ln:line-clean} of \Cref{alg:line}, run with this $h$, visits the required configurations of the
cleanup and has movement $O\bigl(n\sqrt{S_{h}}+h\,S_{h}+n\bigr)$.
\end{lemma}

\begin{proof}
By \Cref{eq:macro}, $P\le S_{h}$. By \Cref{lem:batch}(i) on the grid of $\lceil n/h\rceil^{2}$ product
boxes, the lattice path has $O\bigl(\lceil n/h\rceil\sqrt{P}+\lceil n/h\rceil\bigr)$ steps, each of
movement $O(h)$, so it contributes $O\bigl(n\sqrt{S_{h}}+n\bigr)$. Each occupied $I\times J$ holds at
most $Q_{I,J}$ required configurations, so by \Cref{lem:batch}(i) visiting them and returning takes
movement $O\bigl(h\sqrt{Q_{I,J}}\bigr)$, because $Q_{I,J}\ge 1$. By \Cref{eq:macro}, these visits total
$O(h\,S_{h})$.
\end{proof}

\begin{theorem}
\label{thm:line-ub}
For every fixed $r$, an $r$-round algorithm sorts the line with expected movement
$O_{r}\bigl(n^{\beta_{r}}\bigr)$ on every input, and its output is correct for every outcome of its
coins. For $r\ge 2$ this is \Cref{alg:line}.
\end{theorem}

\begin{proof}
For $r=1$, a serpentine route through $[n]^{2}$ visits every cell, with movement
$O(n^{2})=O(n^{\beta_{1}})$, and it is deterministic. Let $r\ge 2$. Correctness was shown after
\Cref{alg:sampling}. We bound the rounds in turn, writing $L_{t}$ for the movement of round $t$. The
expectations are over the coins.

Round $1$ has movement $O(n^{\beta})$, as noted after \Cref{alg:sampling}. By \Cref{lem:splitter}
with $s=\lceil n/b_{0}\rceil$, its output has squared mass $A$ with $\mathbb{E}[A]\le 3nb_{0}$.

We first bound the dispersion of the family that enters each later round: for $1\le j\le r-1$, the family entering
round $j+1$ satisfies $\mathbb{E}[S_{h_{j}}]\le 6nb_{j-1}$. For $j=1$ this holds because $A$ is the
dispersion for intervals of length $1$, so $S_{h_{1}}\le 2A$. For $j\ge 2$, \Cref{lem:sample} applied
to round $j$ gives $\mathbb{E}[S_{h_{j-1}}]\le 3n/p_{j-1}=3nb_{j-1}$, whatever the family entering round $j$, and
$S_{h_{j}}\le 2S_{h_{j-1}}$ because $h_{j-1}\le h_{j}$.

Now let $1\le j\le r-2$. By \Cref{lem:refine},
\[
  \mathbb{E}[L_{j+1}]=O\bigl(p_{j}n^{2}/h_{j}+p_{j}h_{j}\,nb_{j-1}+n\bigr)
   =O\bigl(p_{j}\,n\sqrt{nb_{j-1}}+n\bigr),
\]
because $\sqrt{n/b_{j-1}}\le h_{j}\le 2\sqrt{n/b_{j-1}}$. The exponent of $p_{j}\,n\sqrt{nb_{j-1}}$
is $\tfrac32+a_{j-1}/2-a_{j}$, which equals $\beta$ by \Cref{eq:bal-ref} with $D=2$. Hence
$\mathbb{E}[L_{j+1}]=O(n^{\beta})$.

For round $r$, let $h=h_{r-1}$. By \Cref{lem:cleanup} and Jensen's inequality for the square root,
\[
  \mathbb{E}[L_{r}]=O\bigl(n\sqrt{ng}+\sqrt{n/g}\cdot ng+n\bigr)=O\bigl(n^{3/2}\sqrt{g}\bigr).
\]
The exponent $\tfrac32+a_{r-2}/2$ equals $\beta$ by \Cref{eq:bal-clean} with $D=2$. Summing over the
$r$ rounds gives $\mathbb{E}[\mathrm{Move}]=O_{r}(n^{\beta})$.
\end{proof}

Together with \Cref{thm:line-exact}, this gives the exact order on the line.

\begin{remark}
\label{rem:selection}
On the line, $\beta_{r}=1+\tfrac12 s_{r}$, where $s_{r}=1+1/(2^{r}-1)$ is the
exponent of the worst-case number of comparisons that a deterministic algorithm needs to find the
maximum of $n$ keys, or to merge two sorted lists, in $r$ rounds~\cite{Valiant75,HH82}. Here $g=n^{1/(2^{r}-1)}$, so the finest scale of \Cref{sec:lb} has
$\Theta(ng)=\Theta(n^{s_{r}})$ co-blocked cells. Unless an earlier round already has expected movement $\Omega(n^{\beta_{r}})$, with
constant probability the final round must visit a constant fraction of them, pairwise at distance
$\Omega\bigl((n/g)^{1/2}\bigr)$, which takes movement $\Omega(n^{\beta_{r}})$.
\end{remark}

\section{Higher Dimensions}
\label{sec:highdim}

This section completes the upper-bound half of \Cref{thm:main}. For $d\ge 2$, the three ranges of $r$
use three kinds of refinement: none for $r\le 2$ (\Cref{subsec:hd-endpoints}), two independent
samples in one round for $r=3$ (\Cref{subsec:hd-r3}), and one sample for each box size of a nested
partition for $r\ge 4$ (\Cref{subsec:interior}). The section ends with the proof of \Cref{thm:main}.

\subsection{At Most Two Rounds}
\label{subsec:hd-endpoints}

\begin{proposition}
\label{prop:hd-endpoints}
Let $d\ge 2$. Then (i) $\Mof{1}{d}(n)=O_{d}(n^{2})$, attained by a deterministic algorithm, and
(ii) $\Mof{2}{d}(n)=O_{d}\bigl(n^{\beta_{2}^{(d)}}\bigr)$, where $\beta_{2}^{(d)}=\tfrac{6d-1}{4d-1}$.
\end{proposition}

\begin{proof}
For $r=1$, a serpentine route through $[m]^{D}$ visits every cell, with movement $O_{d}(n^{2})$, and it
is deterministic.

For $r=2$, we run \Cref{alg:sampling} without refinement rounds, with $b_{0}=n^{a_{0}}$ for the
exponent $a_{0}$ of \Cref{lem:schedule} at $r=2$. Round $1$ has movement $O(n^{\beta})$, where
$\beta=\beta_{2}^{(d)}$. By \Cref{lem:splitter} with $s=\lceil n/b_{0}\rceil$, its output has
squared mass $A$ with $\mathbb{E}[A]\le 3nb_{0}$. Round $2$ has at most $A$ required configurations, and they
are known, so by \Cref{lem:batch}(ii) and Jensen's inequality for the concave function
$t\mapsto t^{1-1/D}$, its expected movement is
$O_{d}\bigl(n^{1/d}(nb_{0})^{1-1/D}+n^{1/d}\bigr)$. The exponent
$\tfrac1d+(1+a_{0})\bigl(1-\tfrac1D\bigr)$ equals $1+a_{0}+(1-a_{0})/D$, because $1/d=2/D$, and this
is $\beta$ by \Cref{eq:bal-clean}.
\end{proof}

\subsection{Two Samples at \texorpdfstring{$r=3$}{r=3}}
\label{subsec:hd-r3}

The cost of the cleanup has two parts. Its route moves along a lattice path through the occupied
product boxes $I\times J$ of two cubes of side $H$, and inside each of them it visits the required configurations, with
movement $O_{d}\bigl(H\,Q_{I,J}^{1-1/D}+H\bigr)$ by \Cref{lem:batch}(i). On the line, $1-1/D=\tfrac12$, so by
\Cref{eq:macro} the dispersion bounds both parts, and one sample of intervals suffices. For $d\ge 2$ the
exponent $1-1/D$ is larger, and we bound a visit by $H\,Q_{I,J}$, whose sum over the boxes is $H$ times the squared mass. Sampling cubes controls the
dispersion, which bounds the lattice path, and sampling keys controls the squared mass, which bounds the visits, so
the refinement round does both.

We write $\mathcal{P}_{H}(G)$ and $S_{H}$ for the partition of $[m]^{d}$ into cubes of side $H$.

\begin{definition}[Cube and Key Samples]
\label{def:twosample}
A \emph{cube sample} is a sample whose partition consists of the cubes of side $H$. A \emph{key
sample} is a sample on the partition into single positions.
\end{definition}

Take $a_{0}>a_{1}$ from \Cref{lem:schedule} at $r=3$, and let $b_{0}=n^{a_{0}}$, $A_{\star}=nb_{0}$,
and $H=\lceil n^{(1-a_{0})/D}\rceil$. Let $p=n^{-a_{1}}$ and
$\sigma=n^{-(a_{0}-a_{1})}$. Both rates lie in $(0,1)$. The rate $p$ serves the tour of the cleanup, since a cube
sample leaves dispersion about $n/p$, and the rate $\sigma$ serves its visits, since a key sample leaves squared mass about
$n/\sigma$. The minimum corner of a cube is its point
with all coordinates smallest. For the family entering round $2$, let $P_{J}$ be the number of cubes
$I$ with $Q_{I,J}>0$.

\begin{algorithm}[t]
\caption{The two-sample algorithm}
\label{alg:twosample}
\begin{algorithmic}[1]
\REQUIRE $d\ge 2$; $H$, $p$, and $\sigma$ as above.
\STATE \textbf{Round $1$ (splitter).} Lines~\ref{ln:split-draw}--\ref{ln:split-state} of
  \Cref{alg:sampling}.
\STATE \textbf{Round $2$ (refinement).} Draw a cube sample of side $H$ and rate $p$ and a key sample
  of rate $\sigma$, independently.
\FOR{each selected cube $J$ with $P_{J}\ge 1$, in the serpentine order of the cubes}
  \label{ln:two-cube-begin}
  \STATE Place the second head at the minimum corner of $J$, and move the first head along the
    lattice path of \Cref{lem:batch}(i) through the cubes $I$ with $Q_{I,J}>0$.
  \STATE On arriving at such a cube $I$, visit the required configurations of the cube sample in $I\times J$
    by \Cref{lem:batch}(i). Then return the second head to the minimum corner of $J$ and the first
    head to its place on the path. \label{ln:two-cube-end}
\ENDFOR
\STATE Visit the required configurations of the key sample by \Cref{lem:batch}(ii). \label{ln:two-key}
\STATE Update the state as in Line~\ref{ln:ref-state} of \Cref{alg:sampling}.
\STATE \textbf{Round $3$ (cleanup).} Move along the lattice path of \Cref{lem:batch}(i) through the
  occupied product boxes on the grid of product boxes of cubes of side $H$. On first arrival at an
  occupied $I\times J$, visit the required configurations in it by \Cref{lem:batch}(i). \label{ln:two-clean}
\STATE Output as in Line~\ref{ln:output} of \Cref{alg:sampling}.
\end{algorithmic}
\end{algorithm}

The next lemma bounds the route of the cube sample for any family.

\begin{lemma}
\label{lem:block}
Let a family of disjoint segments have squared mass $A$, let $1\le H\le n^{1/d}$, and let $p\in(0,1]$.
The route of Lines~\ref{ln:two-cube-begin}--\ref{ln:two-cube-end} of \Cref{alg:twosample} visits the
required configurations of a cube sample of side $H$ and rate $p$, and its expected movement is
\[
  O_{d}\bigl(p\,n^{2/d}H^{-1}A^{1-1/d}+p\,H\,A+n\bigr).
\]
\end{lemma}

\begin{proof}
We bound in turn the moves of the second head between cubes, the paths of the first head, and the
visits inside product boxes. There are at most $(2n^{1/d}/H)^{d}=2^{d}n/H^{d}$ cubes. The second head
moves through them in serpentine order at $O_{d}(H)$ per cube, which totals
$O_{d}(n/H^{d-1})=O_{d}(n)$.

Fix a selected cube $J$ with $P_{J}\ge 1$. By \Cref{lem:batch}(i) on the grid of cubes, the path of
the first head has $O_{d}\bigl((n^{1/d}/H)P_{J}^{1-1/d}\bigr)$ steps, each of movement $O_{d}(H)$.
Here $P_{J}\ge 1$ absorbs the additive term, and also the $O_{d}(n^{1/d})$ movement of the first head
from the previous path to this one. So the first head moves $O_{d}\bigl(n^{1/d}P_{J}^{1-1/d}\bigr)$
for $J$. Each $J$ is selected with probability $p$. Moreover,
$\sum_{J}P_{J}=P\le\sum_{G}|\mathcal{P}_{H}(G)|^{2}\le A$ by \Cref{eq:macro} and
$|\mathcal{P}_{H}(G)|\le|G|$. By the concavity of $t\mapsto t^{1-1/d}$ over the at most
$2^{d}n/H^{d}$ cubes,
\[
  \mathbb{E}\sum_{J\text{ selected}}P_{J}^{1-1/d}
   \ \le\ p\Bigl(\frac{2^{d}n}{H^{d}}\Bigr)^{1/d}A^{1-1/d}
   \ =\ 2p\,n^{1/d}H^{-1}A^{1-1/d},
\]
so the first head moves $O_{d}\bigl(p\,n^{2/d}H^{-1}A^{1-1/d}\bigr)$ in expectation.

Finally, for an occupied $I\times J$ with $J$ selected, at most $Q_{I,J}$ required configurations lie in
$I\times J$, so by \Cref{lem:batch}(i) visiting them and returning takes movement
$O_{d}\bigl(H\,Q_{I,J}^{1-1/D}+H\bigr)=O_{d}(H\,Q_{I,J})$, because $Q_{I,J}$ is a positive integer.
Since $\sum_{I,J}Q_{I,J}=\sum_{G}|G|^{2}=A$ and each $J$ is selected with probability $p$, these
visits have expected total $O_{d}(p\,H\,A)$.
\end{proof}

For a family of squared mass $A$, the key sample has at most $\sigma A$ required configurations in expectation,
because each ordered pair of distinct keys in a common segment is required with probability $\sigma$.
By \Cref{lem:batch}(ii) and Jensen's inequality,
Line~\ref{ln:two-key} of \Cref{alg:twosample} has expected movement
$O_{d}\bigl(n^{1/d}(\sigma A)^{1-1/D}+n^{1/d}\bigr)$.

\begin{theorem}
\label{thm:r3}
For every $d\ge 2$, \Cref{alg:twosample} sorts with expected movement $O_{d}\bigl(n^{\beta_{3}^{(d)}}\bigr)$
on every input, and its output is correct for every outcome of its coins, where
$\beta_{3}^{(d)}=\tfrac{16d^{2}-6d+1}{12d^{2}-6d+1}$. Hence $\Mof{3}{d}(n)=O_{d}\bigl(n^{\beta_{3}^{(d)}}\bigr)$.
\end{theorem}

\begin{proof}
Write $\beta=\beta_{3}^{(d)}$. Correctness was shown after \Cref{alg:sampling}. We bound the rounds in
turn. Round $1$ has movement $O(n^{\beta})$. By \Cref{lem:splitter}, its output has squared mass $A$
with $\mathbb{E}[A]\le 3A_{\star}$.

For the cube sample in round $2$, we take expectations in \Cref{lem:block}, using Jensen's inequality
for $t\mapsto t^{1-1/d}$. The value of $H$ makes the two main terms equal: since $H=\Theta\bigl(n^{(1-a_{0})/D}\bigr)$, both are
\[
  O_{d}\bigl(p\,n^{2/d}H^{-1}A_{\star}^{1-1/d}+p\,H\,A_{\star}\bigr)
   =O_{d}\bigl(n^{-a_{1}+1+a_{0}+(1-a_{0})/D}\bigr)=O_{d}(n^{\beta}),
\]
where the last step is \Cref{eq:bal-ref} for $j=1$. For the key sample, $\sigma A_{\star}=n^{1+a_{1}}$,
so its expected movement is $O_{d}\bigl(n^{1/d+(1+a_{1})(1-1/D)}+n^{1/d}\bigr)$. The exponent equals
$1+a_{1}+(1-a_{1})/D$, which is $\beta$ by \Cref{eq:bal-clean}.

For round $3$, the output of round $2$ is obtained from the output of the cube sample by splitting
segments, so \Cref{lem:sample} gives $\mathbb{E}[S_{H}]\le 3n/p$. By \Cref{eq:macro}, $P\le S_{H}$. By
\Cref{lem:batch}(i) on the grid of product boxes, the lattice path has
$O_{d}\bigl((n^{1/d}/H)(P^{1-1/D}+1)\bigr)$ steps, each of movement $O_{d}(H)$. By Jensen's
inequality, the path has expected movement $O_{d}\bigl(n^{1/d}(n/p)^{1-1/D}+n^{1/d}\bigr)$. Since
$n/p=n^{1+a_{1}}$, this is $O_{d}(n^{\beta})$ by the computation for the key sample. By
\Cref{lem:batch}(i) and $Q_{I,J}\ge 1$, the visits inside the occupied product boxes take movement
$O_{d}\bigl(H\sum_{I,J}Q_{I,J}\bigr)$, and $\sum_{I,J}Q_{I,J}$ is the squared mass of the output of round $2$. This
output is obtained from the output of the key sample by splitting segments, so \Cref{lem:sample} for the partition
into single positions shows that its squared mass has expectation at most $3n/\sigma$. Hence the visits have expected
movement $O_{d}(Hn/\sigma)=O_{d}\bigl(n^{1+(1-a_{0})/D+a_{0}-a_{1}}\bigr)=O_{d}(n^{\beta})$, where the last step is
\Cref{eq:bal-ref} for $j=1$.

Summing over the three rounds gives $\mathbb{E}[\mathrm{Move}]=O_{d}(n^{\beta})$.
\end{proof}

\subsection{Multiscale Refinement for \texorpdfstring{$r\ge 4$}{r >= 4}}
\label{subsec:interior}

At $r=3$ the refinement round is followed by the cleanup, whose cost is controlled by one dispersion and
the squared mass. For $r\ge 4$ a refinement round can be followed by another one, whose route moves
at every box size (\Cref{lem:ms-route}), so its cost depends on the dispersion of its input at all these sizes. So the analysis keeps an upper bound on the expected dispersion for each box size, and
the rates of each round are chosen in advance from this bound.

In this subsection $d\ge 2$ and $r\ge 4$ are fixed and $\beta=\beta_{r}^{(d)}$. A scale here means a
spatial scale: $x$ and $y$ denote volume exponents of boxes, positions are written $u$ and $v$, and
the rank scales enter only through the exponents $a_{j}$. Let $c=1/d$, and recall that $\delta=\beta-1$
and $a_{0}=1-\delta$.

\begin{definition}[Scales and Samples at a Scale]
\label{def:multiscale}
\begin{enumerate}[label=\textup{(\arabic*)}]
  \item Halving every axis $i$ times, for $0\le i\le\lfloor\log_{2}m\rfloor$, partitions $[m]^{d}$
        into $2^{id}$ boxes with sides $\lfloor m/2^{i}\rfloor$ and $\lceil m/2^{i}\rceil$. Such a box
        is a \emph{scale-$x$ box}, where $n^{x}=(m/2^{i})^{d}$, and $x$ is its \emph{spatial scale}.
        We call these $x$ \emph{dyadic}, and also $x=0$, whose scale-$0$ boxes are the single
        positions. These partitions are nested, and a box contains at most $2^{d}$ boxes of the next
        finer partition. There are $B_{x}=n^{1-x}$ scale-$x$ boxes, with sides at most $2n^{x/d}$, and
        there are $O(\log n)$ dyadic $x$. We write $\mathcal{P}_{x}(G)$ and $S_{x}$ for the partition into
        scale-$x$ boxes. Thus $S_{0}$ is the squared mass, and $S_{x}\le\min\{S_{0},nB_{x}\}$ because
        $|\mathcal{P}_{x}(G)|\le\min\{|G|,B_{x}\}$.
  \item A \emph{sample at scale $y$} is a sample on the partition into scale-$y$ boxes. Its required
        configurations are the set $W_{y}$ of pairs $(u,v)$ with $u\ne v$ and $v$ a pivot of this
        sample in the segment of $u$. For a set $W$ of configurations, $N_{x}(W)$ is the number of
        product boxes of two scale-$x$ boxes that contain a configuration of $W$. Thus
        $N_{x}(W)\le\min\{|W|,B_{x}^{2}\}$.
\end{enumerate}
\end{definition}

\begin{lemma}
\label{lem:ms-route}
A set $W$ of configurations is visited by a route that starts at the current configuration and has
movement $O_{d}\bigl(n^{1/d}+\sum_{x}n^{cx}N_{x}(W)\bigr)$, where the sum ranges over the dyadic
$x$. The route can be computed from $W$ in polynomial time.
\end{lemma}

\begin{proof}
The product boxes that contain a configuration of $W$ form a tree under inclusion, with the root
$[m]^{D}$. A box has at most $2^{D}$ children, and the leaves are the single configurations of $W$.
Mark one configuration of $W$ in each node, and traverse the tree in depth-first order, moving between
marked configurations. Each edge is crossed twice. A crossing into or out of a child of a scale-$x$
node moves $O_{d}(n^{cx})$, because both marked configurations lie in this node. So the traversal
visits every configuration of $W$ and moves $O_{d}\bigl(\sum_{x}n^{cx}N_{x}(W)\bigr)$, and it moves
$O_{d}(n^{1/d})$ from the current configuration to the mark of the root.
\end{proof}

For the required configurations of a sample, the next lemma bounds $N_{x}$ at the coarse scales.

\begin{lemma}
\label{lem:ms-occ}
Fix a family of disjoint segments and draw a sample at scale $y$ with rate $p$. For every dyadic $x$,
$\mathbb{E}\bigl[N_{x}(W_{y})\bigr]\le p\,S_{x}$.
\end{lemma}

\begin{proof}
A configuration of $W_{y}$ consists of a key of a segment $G$ and a pivot of $G$. So a segment $G$
contributes to $N_{x}(W_{y})$ only through product boxes $I\times J$ such that $I$ meets $G$ and $J$
contains a pivot of $G$. Let $t_{x}(G)$ be the number of scale-$x$ boxes that contain a pivot of $G$.
There are at most $|\mathcal{P}_{x}(G)|\,t_{x}(G)$ such product boxes. If $x\le y$, a scale-$x$ box that meets $G$ contains a pivot of $G$ exactly when the scale-$y$ box
containing it is selected, so $\mathbb{E}[t_{x}(G)]=p\,|\mathcal{P}_{x}(G)|\le p\,|G|$. If $x\ge y$,
a scale-$x$ box that contains a pivot of $G$ contains a selected scale-$y$ box that meets $G$, so
$\mathbb{E}[t_{x}(G)]\le p\,|\mathcal{P}_{y}(G)|\le p\,|G|$. In both cases
$\mathbb{E}\bigl[|\mathcal{P}_{x}(G)|\,t_{x}(G)\bigr]\le p\,|G|\,|\mathcal{P}_{x}(G)|$, and summing over
the segments gives the bound.
\end{proof}

The rate alone gives one more bound, for every family. A product box $I\times J$ contains a
configuration of $W_{y}$ only if $J$ contains a pivot. If $x\le y$, this requires the scale-$y$ box
containing $J$ to be selected, so $\mathbb{E}[N_{x}(W_{y})]\le p\,B_{x}^{2}$. If $x\ge y$, it requires
$J$ to contain a selected scale-$y$ box, and $p\,B_{y}$ scale-$y$ boxes are selected in expectation,
so $\mathbb{E}[N_{x}(W_{y})]\le p\,B_{x}B_{y}$. In both cases,
\begin{equation}
  \label{eq:rate}
  \mathbb{E}[N_{x}(W_{y})]\ \le\ p\,n^{2-x-\min\{x,y\}}.
\end{equation}

Only the scales $x\le x_{\star}$, where $x_{\star}=a_{0}/(2-c)$, require control of the segments. For dyadic $x\ge x_{\star}$,
$n^{cx}B_{x}^{2}=n^{2-(2-c)x}\le n^{2-a_{0}}=n^{\beta}$, so each of these scales contributes at most
$n^{\beta}$ to the bound of \Cref{lem:ms-route} for any set of configurations.

\begin{definition}[Profiles]
\label{def:profile}
A \emph{profile} of a random family of segments is a function $f\colon[0,x_{\star}]\to\mathbb{R}$
with $\mathbb{E}[S_{x}]\le 3n^{1+f(x)}$ at every dyadic $x\le x_{\star}$. For a function $f$ on
$[0,x_{\star}]$, let $\varphi_{f}(x)=f(x)+cx-\delta$ and
$\psi_{y}(x)=a_{0}-(1-c)x-\min\{x,y\}$, and let
\[
  (\Phi f)(y)=\sup_{0\le x\le x_{\star}}\ \min\bigl\{\varphi_{f}(x),\,\psi_{y}(x)\bigr\}.
\]
\end{definition}

The functions $\varphi_{f}$ and $\psi_{y}$ give two ways to keep the contribution of a scale $x$ to the
movement of a sample at scale $y$ at $O(n^{\beta})$: through the dispersion, with rate at most $n^{-\varphi_{f}(x)}$, or
through the rate alone, with rate at most $n^{-\psi_{y}(x)}$. So $n^{-(\Phi f)(y)}$ is the largest rate for
which one of the two applies at every $x\le x_{\star}$ (\Cref{lem:round}).

We use two facts about $x_{\star}$, in \Cref{lem:ms-iter} and \Cref{thm:interior}. First,
\begin{equation}
  \label{eq:xstar}
  a_{r-2}=\delta-c\,x_{\star}.
\end{equation}
Indeed, \Cref{lem:schedule} gives $\delta(1-q^{r})=1-q$ and $a_{r-2}=q^{r-1}\delta$, and $c=2(1-q)$
and $2-c=2q$, so
\[
  \delta-c\,x_{\star}
   =\delta-\frac{c\,(1-\delta)}{2-c}
   =\frac{\delta-(1-q)}{q}
   =\frac{\delta\,q^{r}}{q}
   =a_{r-2}.
\]
Second, $\delta\le x_{\star}$. This inequality is equivalent to $(1-q^{r})/(1-q)\ge 1+2q$, which holds
for $r\ge 4$ because $q\ge\tfrac34$ gives $q^{2}+q^{3}\ge q$.

Let $f_{0}(x)=\min\{a_{0},1-x\}$ and $f_{j}=\Phi f_{j-1}$ for $1\le j\le r-2$. Thus $f_{0}$ is the bound that the
splitters and the trivial bound $S_{x}\le n^{2-x}$ give, and $f_{j}$ is the bound that we keep after round $j+1$. \Cref{alg:multiscale} is
the instance of \Cref{alg:sampling} that draws one sample per dyadic scale.

\begin{algorithm}[t]
\caption{The multiscale algorithm}
\label{alg:multiscale}
\begin{algorithmic}[1]
\REQUIRE $d\ge 2$; $r\ge 4$; the functions $f_{0},\dots,f_{r-2}$ as above.
\STATE \textbf{Round $1$ (splitter).} Lines~\ref{ln:split-draw}--\ref{ln:split-state} of
  \Cref{alg:sampling}.
\FOR{$j=1,\dots,r-2$}
  \STATE \textbf{Round $j+1$ (refinement).} For every dyadic $y\le x_{\star}$, draw a sample at scale
    $y$ with rate $n^{-f_{j}(y)}$, independently. \label{ln:ms-draw}
  \STATE For every such $y$, visit $W_{y}$ by the route of \Cref{lem:ms-route}, and join these routes end
    to end. \label{ln:ms-route}
  \STATE Update the state as in Line~\ref{ln:ref-state} of \Cref{alg:sampling}.
\ENDFOR
\STATE \textbf{Round $r$ (cleanup).} Visit the required configurations by the route of \Cref{lem:ms-route}.
\STATE Output as in Line~\ref{ln:output} of \Cref{alg:sampling}.
\end{algorithmic}
\end{algorithm}

The next lemma is the inductive step of the analysis: it bounds the movement of one refinement round
and the profile of its output.

\begin{lemma}
\label{lem:round}
Let $f$ be a profile of the family entering a refinement round, and let $(\Phi f)(y)\ge 0$ at every
dyadic $y\le x_{\star}$. Suppose that the round draws, for every such $y$, a sample at scale $y$ with
rate $n^{-(\Phi f)(y)}$, independently, and visits $W_{y}$ by the route of \Cref{lem:ms-route}. Then the
round has expected movement $O_{d}(n^{\beta}\log^{2}n)$, and $\Phi f$ is a profile of its output.
\end{lemma}

\begin{proof}
We bound the contribution of each scale to each sample, and then the profile of the output. Fix a dyadic
$y\le x_{\star}$, let $p=n^{-(\Phi f)(y)}$, and fix a dyadic $x$. We show that
$n^{cx}\,\mathbb{E}[N_{x}(W_{y})]\le 3n^{\beta}$. If $x\ge x_{\star}$, this follows from
$N_{x}(W_{y})\le B_{x}^{2}$ and $n^{cx}B_{x}^{2}\le n^{\beta}$. Let $x\le x_{\star}$. By the definition of
$\Phi$, $p\le n^{-\min\{\varphi_{f}(x),\psi_{y}(x)\}}$, so $p\le n^{-\varphi_{f}(x)}$ or
$p\le n^{-\psi_{y}(x)}$. In the first case, \Cref{lem:ms-occ} and the profile give
\[
  n^{cx}\,\mathbb{E}[N_{x}(W_{y})]\ \le\ n^{cx}\,p\,\mathbb{E}[S_{x}]
   \ \le\ 3\,n^{cx-f(x)-cx+\delta+1+f(x)}\ =\ 3n^{\beta}.
\]
In the second case, \Cref{eq:rate} gives
\[
  n^{cx}\,\mathbb{E}[N_{x}(W_{y})]\ \le\ n^{cx-a_{0}+(1-c)x+\min\{x,y\}+2-x-\min\{x,y\}}\ =\ n^{2-a_{0}}\ =\ n^{\beta}.
\]
By \Cref{lem:ms-route}, the route of this sample has expected movement
$O_{d}(n^{1/d}+n^{\beta}\log n)$, since there are $O(\log n)$ dyadic $x$. There are $O(\log n)$ samples,
which gives the bound on the movement.

For the output, \Cref{lem:sample} shows that the refined segments of the sample at scale $y$ satisfy
$\mathbb{E}[S_{y}]\le 3n/p=3n^{1+(\Phi f)(y)}$, whatever the entering family. The output of the round
is obtained from them by splitting segments, so it satisfies the same bound.
\end{proof}

The cleanup needs the value of $f_{r-2}$ at $x_{\star}$. The next lemma gives it, together with the monotonicity that
carries it to smaller $x$.

\begin{lemma}
\label{lem:ms-iter}
For $1\le j\le r-2$, the function $f_{j}$ is nonincreasing, $\varphi_{f_{j}}$ is nondecreasing, and
$f_{j}=a_{j}$ on $[\xi_{j},x_{\star}]$, where $\xi_{j}=(a_{0}-a_{j})/(2-c)$. Consequently $f_{j}\ge a_{j}$.
\end{lemma}

\begin{proof}
We use induction on $j$. For $j=1$, recall that $\delta\le x_{\star}$. On $[0,\delta]$ we have $f_{0}=a_{0}$, so
$\varphi_{f_{0}}(x)=1-2\delta+cx$ increases. On $[\delta,x_{\star}]$ we have $f_{0}(x)=1-x$, so
$\varphi_{f_{0}}(x)=a_{0}-(1-c)x\ge\psi_{y}(x)$, and the minimum in $\Phi$ is $\psi_{y}$, which
decreases. So the supremum is attained on $[0,\delta]$ where $\varphi_{f_{0}}$ and $\psi_{y}$ cross. For
$y\ge\delta/2$ they cross at $x=\delta/2$, and for $y\le\delta/2$ they cross at $x=\delta-y$. This gives
\[
  f_{1}(y)=
  \begin{cases}
    1-(2-c)\delta-cy & \text{for }y\le\delta/2,\\
    1-2\delta+c\delta/2 & \text{for }y\ge\delta/2.
  \end{cases}
\]
Since $q=1-c/2$, the second value equals $q\,a_{0}+c/2-\delta=a_{1}$, and hence
$\xi_{1}=\delta/2$. The formula shows that $f_{1}$ is nonincreasing and $\varphi_{f_{1}}$ is
nondecreasing.

For the step, let $f=f_{j}$, $a=a_{j}$, and $\xi=\xi_{j}$, with $j\le r-3$. Since $f$ is nonincreasing
and $\varphi_{f}$ is nondecreasing, $\varphi_{f}$ is $c$-Lipschitz. The function $\Phi f$ is
nonincreasing because $\psi_{y}$ decreases in $y$. To see that $\varphi_{\Phi f}$ is nondecreasing, let
$z>y$, let $\Delta=z-y$, and for $x\in[0,x_{\star}]$ let $x'=\max\{x-\Delta,0\}$. Then
$\varphi_{f}(x')\ge\varphi_{f}(x)-c\Delta$. Moreover $\psi_{z}(x')\ge\psi_{y}(x)-c\Delta$: for
$x<\Delta$ we have $\psi_{z}(0)=a_{0}\ge\psi_{y}(x)$, and for $x\ge\Delta$ this follows from
$\min\{x-\Delta,z\}\le\min\{x,y\}+\Delta$. Hence $(\Phi f)(z)\ge(\Phi f)(y)-c\Delta$, which is the
claim.

For the value on the interval, let $a'=q\,a+c/2-\delta$ and $\xi'=(a_{0}-a')/(2-c)$. Since $c/2=1/D$,
\Cref{eq:bal-ref} gives $a'=a_{j+1}$, so $a>a'>0$ and $\xi<\xi'<x_{\star}$. Thus $f(\xi')=a$, and a
direct computation from $a_{0}=1-\delta$ and $q+c/2=1$ gives $a+c\xi'-\delta=a'$. Hence, for $y\ge\xi'$,
\[
  \varphi_{f}(\xi')=a'=a_{0}-(2-c)\xi'=\psi_{y}(\xi').
\]
Since $\varphi_{f}\le a'$ to the left of $\xi'$ and $\psi_{y}\le a'$ to the right of it, we get
$(\Phi f)(y)=a'$ for $y\in[\xi',x_{\star}]$.

Finally, $f_{j}\ge f_{j}(x_{\star})=a_{j}$ because $f_{j}$ is nonincreasing.
\end{proof}

Since $f_{j}\ge a_{j}>0$, the rates of \Cref{alg:multiscale} lie in $(0,1)$, and \Cref{lem:round}
applies to round $j+1$ with $f=f_{j-1}$.

\begin{theorem}
\label{thm:interior}
For every fixed $d\ge 2$ and $r\ge 4$, \Cref{alg:multiscale} sorts with expected movement
$O_{d,r}\bigl(n^{\beta_{r}^{(d)}}\log^{2}n\bigr)$ on every input, and its output is correct for every
outcome of its coins. Hence $\Mof{r}{d}(n)=O_{d,r}\bigl(n^{\beta_{r}^{(d)}}\log^{2}n\bigr)$.
\end{theorem}

\begin{proof}
Correctness was shown after \Cref{alg:sampling}. We bound the rounds in turn. Round $1$ has movement
$O(n^{\beta})$. By \Cref{lem:splitter} with $s=\lceil n/b_{0}\rceil$, its output satisfies
$\mathbb{E}[S_{0}]\le 3nb_{0}=3n^{1+a_{0}}$. Since $S_{x}\le S_{0}$ and $S_{x}\le n^{2-x}$, the function
$f_{0}$ is a profile of this family.

By \Cref{lem:round,lem:ms-iter}, for $1\le j\le r-2$, round $j+1$ has expected movement
$O_{d}(n^{\beta}\log^{2}n)$, and $f_{j}$ is a profile of its output.

It remains to bound the cleanup, whose required configurations we call $W$. At a dyadic $x$, a product box of
two scale-$x$ boxes contains a configuration of $W$ only if a segment meets both boxes, so
$N_{x}(W)\le\sum_{G}|\mathcal{P}_{x}(G)|^{2}\le S_{x}$. Also $N_{x}(W)\le B_{x}^{2}$. Let $x\le x_{\star}$.
Since $\varphi_{f_{r-2}}$ is nondecreasing and $f_{r-2}(x_{\star})=a_{r-2}$, \Cref{eq:xstar} gives
\[
  \varphi_{f_{r-2}}(x)\ \le\ \varphi_{f_{r-2}}(x_{\star})\ =\ a_{r-2}+c\,x_{\star}-\delta\ =\ 0.
\]
That is, $cx+f_{r-2}(x)\le\delta$, so $n^{cx}\,\mathbb{E}[S_{x}]\le 3n^{cx+1+f_{r-2}(x)}\le 3n^{\beta}$.
For $x\ge x_{\star}$, we have $n^{cx}B_{x}^{2}\le n^{\beta}$. By \Cref{lem:ms-route}, the cleanup has
expected movement $O_{d}(n^{\beta}\log n)$, since there are $O(\log n)$ dyadic $x$.

Summing over the $r$ rounds gives $\mathbb{E}[\mathrm{Move}]=O_{d,r}(n^{\beta}\log^{2}n)$.
\end{proof}

\begin{proof}[Proof of \Cref{thm:main}]
For finite $r$, the lower bound is \Cref{thm:lb}. For the upper bounds, \Cref{thm:line-ub} treats $d=1$. For
$d\ge 2$, \Cref{prop:hd-endpoints} treats $r\le 2$, \Cref{thm:r3} treats $r=3$, and \Cref{thm:interior} treats
$r\ge 4$. These algorithms are explicit and run in polynomial time. Those for $r=1$ are deterministic routes, and
the others are instances of \Cref{alg:sampling}. This gives $\Mof{r}{d}(n)=\Tht\bigl(n^{\beta_{r}^{(d)}}\bigr)$ for
finite $r$. For $r=\infty$, \Cref{prop:floor,prop:floor-ub} give $\Mof{\infty}{d}(n)=\Theta_{d}\bigl(n^{1+1/(2d)}\bigr)$,
and the algorithm of \Cref{prop:floor-ub} is deterministic, explicit, and runs in polynomial time. The order is exact on the line by
\Cref{thm:line-exact,thm:line-ub} and for $r\le 2$ by \Cref{prop:endpoints} together with
\Cref{thm:line-ub,prop:hd-endpoints}.
\end{proof}

\section{Multiple Heads}
\label{sec:kheads}

So far the model has two heads. With $k\ge 2$ heads on $[m]^{d}$, a configuration is a point of
$([m]^{d})^{k}$, a route is a sequence of configurations at consecutive $\ell_{1}$ distance $1$, and the
outcome at a configuration is the rank order of the distinct keys under the heads. Rounds, correctness,
and the complexity $\Mof{r}{k,d}(n)$ are defined as in \Cref{def:model}, so $\Mof{r}{2,d}=\Mof{r}{d}$. Extra heads
can only help: an algorithm may leave them at the start and ignore what they read, so
$\Mof{r}{k,d}\le\Mof{r}{d}$.

A fixed number of heads nevertheless changes the complexity by at most a constant factor. Two keys are
compared at every configuration that places two heads on them. So all that a $k$-head algorithm learns is read along the
$\binom{k}{2}$ projections of its route to pairs of heads, and each projection, with consecutive repetitions deleted, is a route in the same comparison grid $[m]^{D}$.

\begin{proposition}
\label{prop:kheads}
For every fixed $k\ge 2$ and $d$, and every $r\in\{1,2,\dots\}\cup\{\infty\}$,
$\Mof{r}{k,d}(n)=\Theta_{d,r,k}\bigl(\Mof{r}{d}(n)\bigr)$.
\end{proposition}

\begin{proof}
The upper bound on $\Mof{r}{k,d}$ was shown above. For the lower bound, a two-head algorithm simulates a given $k$-head
algorithm one pair of heads at a time when $r$ is finite, and for $r=\infty$ we bound the movement
directly. Let $r$ be finite. A round of the $k$-head algorithm commits a route $z_{0},\dots,z_{L}$. Let $L^{(i)}$ be the movement of head $i$ on this route, so that
$L=\sum_{i}L^{(i)}$. For a pair $i<j$, let $R_{ij}$ be the sequence of positions of heads $i$ and $j$
along the route, with consecutive repetitions deleted. It is a route in $[m]^{D}$ of movement at most
$L^{(i)}+L^{(j)}$. In one round of its own, the simulator runs the routes $R_{ij}$ in turn. These runs
have total movement at most $\sum_{i<j}(L^{(i)}+L^{(j)})=(k-1)L$, because each head lies in $k-1$
pairs. The moves from one run to the next add $O_{d,k}(n^{1/d})$ per round. The pairwise outcomes
determine the rank order of the keys under the heads at every configuration of the route, so the
simulator knows all outcomes of the round and can compute the next route. Hence
\[
  \Mof{r}{d}(n)\ \le\ (k-1)\,\Mof{r}{k,d}(n)+O_{d,r,k}\bigl(n^{1/d}\bigr).
\]
Since $\Mof{r}{d}(n)=\Omega_{d}\bigl(n^{1+1/(2d)}\bigr)$ by \Cref{prop:floor} and $1+1/(2d)>1/d$, the
additive term is $o\bigl(\Mof{r}{d}(n)\bigr)$, and the proposition follows.

The simulation adds a term in every round, so for $r=\infty$ we argue directly and bound the movement
of an algorithm with any finite number of rounds. By Yao's principle, fix a deterministic $k$-head
algorithm that is correct on every input, and let the input be uniform. Let $\rho=\Theta_{d}(n^{1/(2d)})$ be
the radius of \Cref{lem:scatter} at scale $2$, whose cells join the ranks $2i-1$ and $2i$. With probability at least $7/8$, there is a set
$W\subseteq\mathcal{C}_{2}$ of at least $\tfrac12\lfloor n/2\rfloor$ cells that are pairwise more than
$\rho$ apart. For every cell $\{u,v\}$ of $W$, some configuration of the execution places two heads on
$u$ and $v$. Otherwise, exchanging their two consecutive ranks changes no outcome, so the algorithm gives
the same output on two inputs with different orders. For a pair $i<j$, let $W_{ij}\subseteq W$ be the set
of cells that heads $i$ and $j$ visit together, and let $\widehat{L}^{(i)}$ be the total movement of head
$i$. By \Cref{obs:model}(b), the positions of heads $i$ and $j$ over the whole execution, with
consecutive repetitions and diagonal configurations deleted, form a sequence of cells of $\dist$-length at
most $\widehat{L}^{(i)}+\widehat{L}^{(j)}$. By \Cref{fact:nn}, if $|W_{ij}|\ge 2$, then
$\widehat{L}^{(i)}+\widehat{L}^{(j)}\ge\tfrac12|W_{ij}|\rho$. Summing over the pairs, and noting that the
pairs with $|W_{ij}|\le 1$ cover at most $\binom{k}{2}$ cells, gives
\[
  (k-1)\,\mathrm{Move}\ =\ \sum_{i<j}\bigl(\widehat{L}^{(i)}+\widehat{L}^{(j)}\bigr)
   \ \ge\ \frac{\rho}{2}\Bigl(\frac12\Bigl\lfloor\frac{n}{2}\Bigr\rfloor-\binom{k}{2}\Bigr)
\]
on this event. Hence $\mathbb{E}[\mathrm{Move}]=\Omega_{d,k}\bigl(n^{1+1/(2d)}\bigr)$, with a constant
that does not depend on the number of rounds, so
$\Mof{\infty}{k,d}(n)=\Omega_{d,k}\bigl(n^{1+1/(2d)}\bigr)$. With \Cref{prop:floor-ub}, this is
$\Omega_{d,k}\bigl(\Mof{\infty}{d}(n)\bigr)$.
\end{proof}

\section{Discussion}
\label{sec:discussion}

\Cref{thm:main} determines the movement exponent for all $d$ and $r$. The order is exact on the line, for $r\le 2$, and for
$r=\infty$. In the interior $d\ge 2$, $3\le r<\infty$, the two bounds differ by a factor $O_{d,r}(\log^{3}n)$, and this gap is the one
question \Cref{thm:main} leaves open. The lower bound loses a factor $\log n$ throughout the interior, and the upper bound loses a
factor $\log^{2}n$ for $r\ge 4$ and nothing at $r=3$. We conjecture that both losses can be removed.

\begin{conjecture}
\label{conj:interior}
For every $d\ge 2$ and $3\le r<\infty$, $\Mof{r}{d}(n)=\Theta_{d,r}\bigl(n^{\beta_{r}^{(d)}}\bigr)$.
\end{conjecture}

\section*{AI Methodology Statement}
The author used OpenAI's ChatGPT in Codex and Anthropic's Claude in Claude Code to help organize and polish the
exposition of this paper, based on the author's own drafts. The same tools assisted in surveying the related
literature, in understanding the current state of research in this area and its gaps, and in flagging possible
errors in draft proofs. The author formulated and proved all mathematical claims in this paper, and checked,
revised, and finalized every statement, proof, and reference. The author takes full responsibility for the
content of the submission.

\pagebreak
\bibliographystyle{alpha}
\bibliography{references}

\end{document}